\documentclass[jcp,twocolumn,english,superscriptaddress]{revtex4-1}

\usepackage[T1]{fontenc}
\usepackage[latin9]{inputenc}
\usepackage{color}
\usepackage{babel}
\usepackage{latexsym}
\usepackage{float}
\usepackage{amsmath}
\usepackage{amsfonts}
\usepackage{graphicx}
\usepackage{braket}
\usepackage{times}   
\usepackage{esint}
\usepackage{subfigure}
\usepackage{verbatim}
\usepackage{qcircuit}
\allowdisplaybreaks
\usepackage{appendix}
\usepackage[unicode=true,pdfusetitle,
 bookmarks=false,colorlinks=true,citecolor=blue,urlcolor=blue,linkcolor=red]{hyperref}

\makeatletter

\@ifundefined{textcolor}{}
{%
 \definecolor{BLACK}{gray}{0}
 \definecolor{WHITE}{gray}{1}
 \definecolor{RED}{rgb}{1,0,0}
 \definecolor{GREEN}{rgb}{0,1,0}
 \definecolor{BLUE}{rgb}{0,0,1}
 \definecolor{CYAN}{cmyk}{1,0,0,0}
 \definecolor{MAGENTA}{cmyk}{0,1,0,0}
 \definecolor{YELLOW}{cmyk}{0,0,1,0}
}

\@ifundefined{date}{}{\date{}}
\AtBeginDocument{
  
}
\makeatother

\newcommand{\SAVE}[1]{}

\begin{document}
\renewcommand\abstractname{}

\title{Real time evolution for ultracompact Hamiltonian eigenstates on quantum hardware}
\author{Katherine Klymko}
\email{kklymko@lbl.gov}
\affiliation{Computational Research Division, Lawrence Berkeley National Laboratory, Berkeley, California 94720, USA}
\affiliation{NERSC, Lawrence Berkeley National Laboratory, Berkeley, California 94720, USA}
\author{Carlos Mejuto-Zaera}
\email{carlos_mejutozaera@berkeley.edu}
\affiliation{Computational Research Division, Lawrence Berkeley National Laboratory, Berkeley, California 94720, USA}
\affiliation{Department of Chemistry, University of California, Berkeley, California 94720, USA}
\author{Stephen J. Cotton}
\affiliation{Quantum Artificial Intelligence Laboratory (QuAIL), Exploration Technology Directorate, NASA Ames Research Center, Moffett Field, CA 94035, USA}
\affiliation{KBR, 601 Jefferson St., Houston, TX 77002, USA}
\author{Filip Wudarski}
\affiliation{Quantum Artificial Intelligence Laboratory (QuAIL), Exploration Technology Directorate, NASA Ames Research Center, Moffett Field, CA 94035, USA}
\affiliation{USRA Research Institute for Advanced Computer Science, Mountain View, California 94043, USA}

\author{Miroslav Urbanek}
\affiliation{Computational Research Division, Lawrence Berkeley National Laboratory, Berkeley, California 94720, USA}
\author{Diptarka Hait}
\affiliation{Department of Chemistry, University of California, Berkeley, California 94720, USA}
\affiliation{Chemical Sciences Division, Lawrence Berkeley National Laboratory, Berkeley, California 94720, USA}
\author{Martin Head-Gordon}
\affiliation{Department of Chemistry, University of California, Berkeley, California 94720, USA}
\affiliation{Chemical Sciences Division, Lawrence Berkeley National Laboratory, Berkeley, California 94720, USA}
\author{K. Birgitta Whaley}
\affiliation{Department of Chemistry, University of California, Berkeley, California 94720, USA}
\author{Jonathan Moussa}
\affiliation{ Molecular Sciences Software Institute, Blacksburg, VA 24060, USA}
\author{Nathan Wiebe}
\affiliation{Department of Computer Science, University of Toronto, Canada}
\author{Wibe A. de Jong}
\email{wadejong@lbl.gov}
\affiliation{Computational Research Division, Lawrence Berkeley National Laboratory, Berkeley, California 94720, USA}
\author{Norm M. Tubman}
\email{norman.m.tubman@nasa.gov}
\affiliation{Quantum Artificial Intelligence Laboratory (QuAIL), Exploration Technology Directorate, NASA Ames Research Center, Moffett Field, CA 94035, USA}
\date{\today}

\begin{abstract} 

In this work we present a detailed analysis of variational quantum phase estimation (VQPE), a method based on real-time evolution for ground and excited state estimation on near-term hardware.
We derive the theoretical ground on which the approach stands, and demonstrate that it provides one of the most compact variational expansions to date for solving strongly correlated Hamiltonians.
At the center of VQPE lies a set of equations, with a simple geometrical interpretation, which provides conditions for the time evolution grid in order to decouple eigenstates out of the set of time evolved expansion states, and connects the method to the classical filter diagonalization algorithm. 
Further, we introduce what we call the unitary formulation of VQPE, in which the number of matrix elements that need to be measured scales linearly with the number of expansion states, and we provide an analysis of the effects of noise which substantially improves previous considerations.
The unitary formulation allows for a direct comparison to iterative phase estimation.  
Our results mark VQPE as both a natural and highly efficient quantum algorithm for ground and excited state calculations of general many-body systems.
We demonstrate a hardware implementation of VQPE for the transverse field Ising model.
Further, we illustrate its power on a paradigmatic example of strong correlation (Cr$_2$ in the def2-SVP basis set), and show that it is possible to reach chemical accuracy with as few as $\sim$50 timesteps. 
\end{abstract}
\maketitle

\newpage

\section{Introduction}

In fulfilling the promise of quantum computation~\cite{Arute2019,Arute2020a,Arute2020b} 
and enabling the exact solution of many-body quantum systems, numerous algorithms of different resource requirements have been proposed~\cite{cerezo2020variational, Motta2020b,kirby2020vqe,nielson2006quantum}, which require quantum and classical resources of different complexity. 
Many of these algorithms are focused on efficient eigenvalue extraction, important for solving problems in chemistry~\cite{mcardle2020quantum}, physics~\cite{smith2019simulating}, and materials science~\cite{doi:10.1021/acs.chemrev.9b00829} and limited classically by the exponential scaling of Hilbert space with system size.
Though immense progress has been made in the development of quantum algorithms for eigenvalue estimation, the resource requirements remain prohibitively high with regards to noisy  intermediate-scale  quantum (NISQ) hardware~\cite{doi:10.1021/acs.chemrev.8b00803, doi:10.1021/acs.chemrev.9b00829}.

For example, quantum phase estimation (QPE)~\cite{abrams1997simulation,abrams1999quantum,kitaev1997quantum} is considered an algorithm that will need considerable quantum resources to run, but will ultimately be a highly accurate approach to quantum simulation. Adiabatic state preparation~\cite{aspuru2005simulated, hauke2020perspectives} allows the ground state of a particular Hamiltonian to be reached by preparing an initial ground state of a simpler system and slowing changing the Hamiltonian to the desired system, requiring long coherence times and low gate errors.
Particularly in the current era of NISQ quantum computers, the variational quantum eigensolver (VQE) framework~\cite{McClean_2016}, and its non-orthogonal variant NOVQE~\cite{huggins2019non}, are promising approaches for the exact solution of many-body quantum systems. 
However, the common formulation of this family of methods relies on the solution of a highly complex variational optimization problem on classical computers, which remains an open challenge~\cite{mcclean2018barren,wierichs2020avoiding}.

The methods described above are generally used to solve the time-independent Schr\"{o}dinger equation to determine the Hamiltonian eigenvalues and eigenstates.
However, time evolution is a more natural operation on a quantum computer and thus simulation of the time-dependent Schr{\"o}dinger equation is a more ideal framework to implement.  
Given the intrinsically quantum-mechanical relation between the time and energy domains~\cite{Tannor2007}, a different family of quantum algorithms focuses on using a time-dependent perspective to solve time-independent problems~\cite{Parrish2019b,Stair2020}.
These methods propose a linear wave function Ansatz expanded in time evolved states and solve the thereby defined generalized eigenvalue equation classically, while the Hamiltonian and overlap matrix elements are measured on quantum hardware.
Exploiting this, quantum computers hold unique potential to outperform their classical counterparts with algorithms based on real-time evolution.
In this work we focus on that advantage and use real-time evolution to generate a basis of states to extract Hamiltonian eigenvalues.

Using real time evolution to generate a basis of states to solve a Hamiltonian is not a new idea~\cite{Neuhauser1990,Neuhauser1994,mandelshtam2001fdm}, but it is not widely used in classical simulation  due to the computational limitations of simulating real time evolution.  
Approximate imaginary time evolution~\cite{ceperley1995path, becca2017quantum} and Krylov diagonalization methods~\cite{Koch2011} are far more widely used in classical simulation, and the intuition behind such approaches is simple to understand: Each new state generated in these approaches has a larger overlap with the true ground state.  
This of course is not how real time evolution works,
as the expectation value of the energy remains constant, and one never gets closer to the ground state during the evolution.   
However, the states that can be generated through real time evolution in fact do provide a basis from which one can extract ground and excited states. 
This is not only highly efficient, but in some cases may be faster than other quantum methods that use time evolution, such as QPE.    
Thus the main goal in this work is to develop an approach for computing ground and excited states, using states generated by real time evolution, that is as fast and efficient as possible.
With this in mind, we analyse the theoretical underpinnings of a class of algorithms we term variational quantum phase estimation (VQPE) which is based on real time expansion methods~\cite{Parrish2019b,Stair2020}. 
We use the term VQPE because of its relationship to both QPE and VQE, as we detail below (see sections~\ref{ss:landscape} and~\ref{ss:vqpeVSqpe}).

Our VQPE algorithm and analysis goes beyond previous proposals in the following ways: We reduce the number of quantum measurements needed to be \emph{linear} instead of quadratic in the number of expansion states. 
We introduce the phase cancellation conditions, providing the underlining theory for why this approach works and deriving a direct link between time steps and band gaps in the spectrum.
We also analyze the effects of noise on our convergence properties, providing significantly improved intuition for ill-conditioning of VQPE methods.
Further, we demonstrate the method classically for several weakly and moderately correlated molecules as well as a strongly correlated transition metal dimer, Cr$_2$.
We show that for all the systems, regardless of the level of electronic correlation, less than 50 real-time evolved expansion states are needed to reach agreement within chemical accuracy for ground state energies obtained with state-of-the-art classical methods requiring $O(10^6)$ variational parameters~\cite{Tubman2016}.
Additionally, we describe and implement the algorithm on quantum hardware for the transverse field Ising model.
Since real-time evolution is natural to implement on quantum hardware, this approach holds immense promise for NISQ implementation.

The paper is structured as follows: in Section~\ref{section:theory}, we analyze the theoretical structure of VQPE. 
This starts with a brief overview and intuition of existing methods corresponding to the VQPE family (\ref{ss:landscape},~\ref{ss:intuiton}). 
Subsequently, we present a novel examination of the theoretical underpinnings of the method based on the phase cancellation picture in~\ref{ss:pcc} and we propose a procedure for choosing optimal time step sizes.  
With the practical implementation of VQPE on NISQ devices in mind, we present the details of our unitary formulation of VQPE in Subsections~\ref{ss:toeplitz} and ~\ref{ss:unitary}.
We conclude the theory section with a careful analysis of the effects of noise~\ref{ss:noise}, a comparison between VQPE and QPE~\ref{ss:vqpeVSqpe}, and a discussion of the inclusion of other time evolved states into VQPE~\ref{ss:other}.
Section~\ref{section:methods} provides details of the systems studied as well as the classical and quantum simulation methods.
Section~\ref{section:results} summarizes and discusses the results of the simulations.
Concluding remarks are found in Section~\ref{section:conclusion}.

\section{Theory}
\label{section:theory}

\subsection{Landscape of Existing Variational Algorithms}
\label{ss:landscape}

VQE approaches are highly relevant for the NISQ era of quantum computation~\cite{cerezo2020variational}. 
They comprise fairly simple quantum circuit implementations at the price of relying on the solution of a high dimensional, classical optimization problem in the presence of noise.  
Despite the optimization challenge, many of the currently existing examples of actual quantum simulations for many-body physics correspond to implementations of this algorithm. 
The basic premise of VQE relies on the variational approximation: A parametrized wave function Ansatz $\ket{\Phi(\vec{\alpha}_j)}$, where $\vec{\alpha}_j$ are the variational parameters, is chosen such that it can be efficiently implemented on a quantum computer. 
The energy expectation value $E(\vec{\alpha}) = \frac{\braket{\Phi(\vec{\alpha}_j)|H|\Phi(\vec{\alpha}_j)}}{\braket{\Phi(\vec{\alpha}_j)|\Phi(\vec{\alpha}_j)}}$ of this Ansatz is evaluated on quantum hardware, and then the optimization problem $\vec{\nabla}E(\vec{\alpha}) = 0$ is solved on a classical computer.
As in any variational approach, the efficacy of the approximation depends on the flexibility of the Ansatz.
A way to increase this flexibility is to choose a more general ground state estimate, namely as a linear combination of several parametrized expansion states $\ket{\Phi_j(\vec{\alpha}_j)}$. 
The total wave function Ansatz thus becomes

\begin{equation}
	\ket{\Psi(\vec{c},\left\{\vec{\alpha}\right\})} = \sum_j c_j \ket{\Phi_j(\vec{\alpha}_j)},
	\label{eq:NOVQE_ansatz}
\end{equation}
where $c_j$ are the expansion coefficients and $\vec{\alpha}_j$ the optimization parameters of the $j$-th expansion states. Applying the variational principle to the coefficients $\vec{c}$ alone leads to the secular equations~\cite{Atkins2010}

\begin{equation}
	\sum_j H_{i,j} c^I_j = \varepsilon_I \sum_j S_{i,j} c^I_j,
	\label{eq:SecEq}
\end{equation}  
where $\varepsilon_I$ is an estimate for the $I$-th Hamiltonian eigenvalue $E_I$. Equation~\ref{eq:SecEq} is a standard generalized eigenvalue equation, which can be solved classically.  
The Hamiltonian, which we assume to be time independent, and overlap matrix elements are measured on quantum hardware in the ``basis" of expansion states following
\begin{equation}
	\begin{split}
		H_{i,j} &= \braket{\Phi_i(\vec{\alpha}_i)|H|\Phi_j(\vec{\alpha}_j)},\\
		S_{i,j} &= \braket{\Phi_i(\vec{\alpha}_i)|\Phi_j(\vec{\alpha}_j)}.
	\end{split}
	\label{eq:HandS}
\end{equation}

Thus, the expansion coefficients $\vec{c}$ can be determined by classically solving the noisy, generalized eigenvalue problem in Eq.~\eqref{eq:SecEq}. 
The expansion states $\ket{\Phi_j(\vec{\alpha}_j)}$ themselves can be then optimized with a classical minimization method, further improving the energy estimates. 
One example in which this has been used recently is with chemically motivated Ans\"{a}tze, such as unitary coupled cluster expansions~\cite{Huggins2020}, which had nonetheless some difficulties related to the optimization of parameters.  

The optimization of the $\vec{\alpha}_j$ parameters is an open field due to (i) the high dimensional nature of the optimization problem and (ii) the presence of noise, which is necessarily part of any approach on quantum hardware.   
To alleviate these problems, an alternative framework has been recently explored~\cite{Mcclean2017,Parrish2019a, Parrish2019b,Takeshita2020,Urbanek2020,Motta2020b, Yeter2020, Stair2020}, which completely bypasses the need for optimization routines. 
In this family of methods, one does not employ a set of parameterized expansion states $\ket{\Phi_j(\vec{\alpha}_j)}$, but instead generates a set of expansion states $\ket{\Phi_j}$ systematically from one or several reference states $\ket{\Psi_I}$. 
The only variational parameters left are thus the expansion coefficient $\vec{c}$, and consequently the only task to be performed by a classical computer is solving the (noisy) generalized eigenvalue problem in Eq.~\eqref{eq:SecEq}.

The way of generating these expansion states should balance ease of implementation on quantum hardware with creating a flexible expansion set, in a variational sense, to obtain accurate energies. 
While \emph{a priori}, by changing from parametrized $\ket{\Phi_j(\vec{\alpha}_j)}$ to systematically generated $\ket{\Phi_j}$, we are reducing the variational flexibility of our Ansatz, the  expansion state generation can still be performed in such a way that it is natural to both the description of ground and excited states, as well as to the implementation on a quantum computer. 
One such approach is the quantum subspace expansion (QSE) method~\cite{Mcclean2017,Takeshita2020,Urbanek2020}, which, after optimizing a ground state estimate with regular VQE, generates expansion states by applying single excitation operators on top of this VQE reference. A recent variation introduced a more general multi-reference Ansatz, targeting ground and exited states simultaneously at the optimization step~\cite{Parrish2019a}. 

Alternatively, the expansion states can be formed by applying the time evolution operator $U(t) = e^{-i H t}$ to the reference states. 
This is the approach followed in the QLanczos method~\cite{Motta2020b,Yeter2020}, where the time evolution is performed along imaginary time (i.e. $t\rightarrow i\tau$), and the quantum filter diagonalization and quantum Krylov approaches~\cite{Stair2020,Parrish2019b}, where the time evolution is performed along real time. 
In the case of the QLanczos method, it is clear that evolving to large enough imaginary times will provide an expansion set which is well suited to describe the ground state, provided that the reference state is not orthogonal to it. 
In the next subsections, we discuss the formalism behind using a set of the real-time states in the expansion, which we refer to as VQPE. 
We also analyze VQPE's robustness to noise, which is critical to asses its applicability in NISQ devices.

\subsection{VQPE - Basic Intuition}
\label{ss:intuiton}

In the VQPE approach~\cite{Stair2020,Parrish2019b}, the expansion states $\ket{\Phi_{j,I}}$ are generated from the reference states $\ket{\Psi_I}$ through time evolution as
\begin{equation}
	\ket{\Phi_{j,I}} = e^{-i H t_j} \ket{\Psi_I}.
	\label{eq:Tevol}
\end{equation}
If there are $N_R$ reference states, and $N_T$ time steps are considered, this produces a ``basis'' of $N_R(N_T + 1)$ expansion states, where $\ket{\Phi_{0,I}}=\ket{\Psi_I}$. For simplicity, we will consider a single reference state $\ket{\Psi_0}$ for now, and will discuss the use of multiple reference states further below.
Throughout the paper, we set the reduced Planck constant $ \hbar= 1$.

On a first glance, it seems counter-intuitive that the set of expansion states in Eq.~\eqref{eq:Tevol} would improve the ground state estimate given by the reference state $\ket{\Psi_0}$. 
After all, the $\ket{\Phi_{j,0}}$ states all have the same energy expectation value. 
Stair \emph{et.~al.}~\cite{Stair2020} suggest an interpretation based on short time evolution, in which Taylor expanding Eq.~\eqref{eq:Tevol} shows that the expansion states span the same space as the Krylov vectors $H^j\ket{\Psi_0}$, making it equivalent to power methods such as the Lanczos algorithm~\cite{Koch2011}. 
This justifies why a set of expansion states concentrated in a time grid over a short time scale should produce a good variational Ansatz for the ground state. 
Including multiple reference states then should provide for good and stable approximates for the first few excited states as well, in the spirit of the band Lanczos method~\cite{Meyer1989}. 
This interpretation suggests that the VQPE method should work best for short time evolution. 
We want to complement this interpretation with a more general one, not limited to short time steps, although this still remains the most interesting regimes from an implementation perspective.
In particular, the VQPE approach is reminiscent of the computation of the autocorrelation function $g(t) = \braket{\Psi_0|e^{-iHt}|\Psi_0}$, which contains the full spectral information of $H$~\cite{Tannor2007}. 
In $g(t)$, the ground state information is encoded in the long time limit, rather than the short time one, since the ground state evolves with the slowest frequency $\omega_0 = E_0$ in units where $\hbar = 1$. 
Thus, there should be nothing particular about the short time evolution limit. 
Instead we articulate the precise requirements for the implementation that will lead to accurate energy estimates.
This is related to the linear independence of the expansion states which poses a lower bound to the optimal time step size.
We also note that VQPE is closely related to the classical filter diagonalization method~\cite{Neuhauser1990,Neuhauser1994,Wall1995,mandelshtam2001fdm}, as pointed out by Parrish \emph{et.al.}~\cite{Parrish2019b}.

We thus want to depart from the Krylov intuition, and provide a different, hopefully more general heuristic, which we will then rigorously formalize, to understand why VQPE approaches should provide good ground and excited state approximations. 
We begin by  writing out the decomposition of the reference state $\ket{\Psi_0}$ into Hamiltonian eigenstates $\ket{N}$, such that $H\ket{N} = E_N\ket{N}$. This can be written out explicitly as

\begin{equation}
    \ket{\Psi_0} = \sum_N \psi^0_N \ket{N},
    \label{eq:RefDecomp}
\end{equation}
where $\psi^0_N = \braket{N|\Psi_0}$ are the coefficients of the reference state $\ket{\Psi_0}$ in the eigenbasis of $H$. We will refer to those Hamiltonian eigenstates $\ket{N}$ for which $\psi^0_N$ is beyond some non-negligible threshold as the \emph{support space} of state $\ket{\Psi_0}$ with respect to (w.r.t.) $H$. This decomposition gives for the expansion states

\begin{equation}
	\ket{\Phi_{j,0}} = e^{-iHt_j}\ket{\Psi_0} = \sum_{N} \psi^0_N e^{-i E_N t_j} \ket{N}.
	\label{eq:PhaseEvol}
\end{equation}

The equation above just states the obvious: each component of $\ket{\Psi_0}$ in the support space w.r.t. $H$ evolves with its own frequency. This, however, makes transparent why the VQPE method can work: Choosing the time grid $\left\{t_j\right\}$ accordingly, it is possible to make linear combinations of the expansion states $\ket{\Phi_{j,0}}$ such that the different phases $e^{-iE_N t_j}$ cancel out targeted components of $\ket{\Psi_0}$ along the support space state $\ket{N}$. In this way it is possible to ``extract'' eigenstates in the support space of $\ket{\Psi_0}$ by including enough expansion states $\ket{\Phi_{j,0}}$. Note that this is not exclusive to the ground state, nor is this limited to short time scales $t_j$. The only requirement is given by the number of eigenstates of $H$ in the support space of $\ket{\Psi_0}$, defining how many time steps are needed for perfect state extraction, and by the energy gaps (i.e. relative frequencies) of those states, which govern the phase cancellation conditions. 

In essence, VQPE allows one to extract ``the most out of the reference state'', in the sense that if there are $Q$ states in its support space, it should be possible to produce $Q$ time evolved states from which to reconstruct the corresponding $Q$ Hamiltonian eigenstates, by solving the secular equation Eq.~\eqref{eq:SecEq}. Of course, this presumes that it is possible to produce $Q$ linearly independent time evolved states, and that our time evolution is noiseless and performed at arbitrary numerical precision. For general reference states, the size of the support space will be too large in general to recover all eigenstates, but a modest amount of these should be enough to approximate the lowest lying energy eigenstates in it.

\subsection{Phase Cancellation Conditions and Relation to Filter-Diagonalization}
\label{ss:pcc}

We now formalize the phase cancellation heuristic on a solid mathematical footing. To this end, we derive a set of equations, the phase cancellation conditions, which set sufficient conditions to exactly extract the Hamiltonian eigenstates from the support space. These conditions embody the intuition in terms of auto-correlation functions described before, and are effectively discrete versions of the main relations at the heart of the classical filter-diagonalization approach~\cite{Neuhauser1990,Neuhauser1994,mandelshtam2001fdm}. 

We consider the overlap matrix $S_{j,k}$ in Eq.~\eqref{eq:HandS} for the expansion states in Eq.~\eqref{eq:PhaseEvol}. It is convenient to write the overlap matrix in operator form in the span of the expansion states $\ket{\Phi_{j,0}}$, and it is easy to verify that

\begin{equation}
    S = \sum_{j=0}^{N_T} \ket{\Phi_{j,0}}\bra{\Phi_{j,0}}.
    \label{eq:Soperator}
\end{equation}

This means that the operator corresponding to the overlap matrix projects onto the span of the expansion states~\footnote{This is easy to confirm, by examining the action of the overlap matrix $S_{j,k}$ on the expansion vectors, which trivially form a basis for their spanned space. 
In this span, we can write $\ket{\Phi_{1,0}} \equiv \begin{pmatrix}1\\0\\ 0 \\ \vdots \\ 0\end{pmatrix}$, $\ket{\Phi_{2,0}} \equiv \begin{pmatrix}0\\1\\ 0 \\ \vdots \\ 0\end{pmatrix}$, and so on, and thus $S \equiv \begin{pmatrix} 1 &  \braket{\Phi_{1,0}|\Phi_{2,0}} & \cdots \\ \braket{\Phi_{2,0}|\Phi_{1,0}} & 1 & \cdots \\ \vdots & \vdots & \vdots \end{pmatrix}$. In this way, we see that $S\ket{\Phi_{k,0}} = \ket{\Phi_{k,0}} + \sum_{j \ne k} \braket{\Phi_{j,0}|\Phi_{k,0}}\ket{\Phi_{j,0}}$, confirming Eq.~\eqref{eq:Soperator}.}. Substituting Eq.~\eqref{eq:PhaseEvol} into the overlap operator gives, after some minor reordering of terms

\begin{equation}
    S = \sum_{N,M}^Q \psi_N^0 \psi_M^{0,*}\left[\sum_{j=0}^{N_T} e^{-i t_j(E_N - E_M)}\right] \ket{N}\bra{M}.
    \label{eq:SopExpansion}
\end{equation}
In the equation above, $Q$ is the number of Hamiltonian eigenstates in the support of $\ket{\Psi_0}$, and we can ignore Hamiltonian eigenstates outside the support space due to their small coefficients $\psi_N^0$. Now, we can define the phase cancellation conditions (PCCs) as

\begin{equation}
    \frac{1}{N_T+1}\sum_{j=0}^{N_T} e^{-i t_j(E_N - E_M)} \stackrel{}{=} \delta_{N,M}. 
    \label{eq:PCC}
\end{equation}
These are the $Q(Q-1)/2$ conditions for the $N_T+1$ time steps in the time grid. Given that the support space is spanned by just $Q$ vectors, it seems that the PCCs impose stricter conditions on the time grid than absolutely necessary to recover the full support space. Still, they embody mathematically the phase cancellation heuristic which we have discussed above. Indeed, the condition in Eq.~\eqref{eq:PCC} enforces the cancellation of the time evolved phase between all Hamiltonian eigenstates in the support of $\ket{\Psi_0}$, and can be represented graphically as a sum of phases in the unit circle. When the phase cancellation conditions are fulfilled, the overlap operator simplifies into a weighted projector into the support of $\ket{\Psi_0}$ w.r.t. $H$, namely

\begin{equation}
    S \stackrel{PCC}{=} \sum_{N}^Q (N_T+1)\left|\psi_N^0\right|^2 \ket{N}\bra{N},
    \label{eq:SopPCC}
\end{equation}
weighted by the coefficients of the reference state $\ket{\Psi_0}$ on the support space, c.f. Eq.~\eqref{eq:RefDecomp}. In this case, the expansion states span exactly the same space as the Hamiltonian eigenstates in the support space, and solving the secular equation~\eqref{eq:SecEq} returns the exact eigenstates and eigenvalues of $H$. The PCCs in Eq.~\eqref{eq:PCC} can be understood as the discrete limit of the eigenstate extraction through Fourier transform of a time evolved state exploited in the classical filter diagonalization literature~\cite{Neuhauser1990}.

\begin{figure*}
    \centering
    \includegraphics[width=\textwidth]{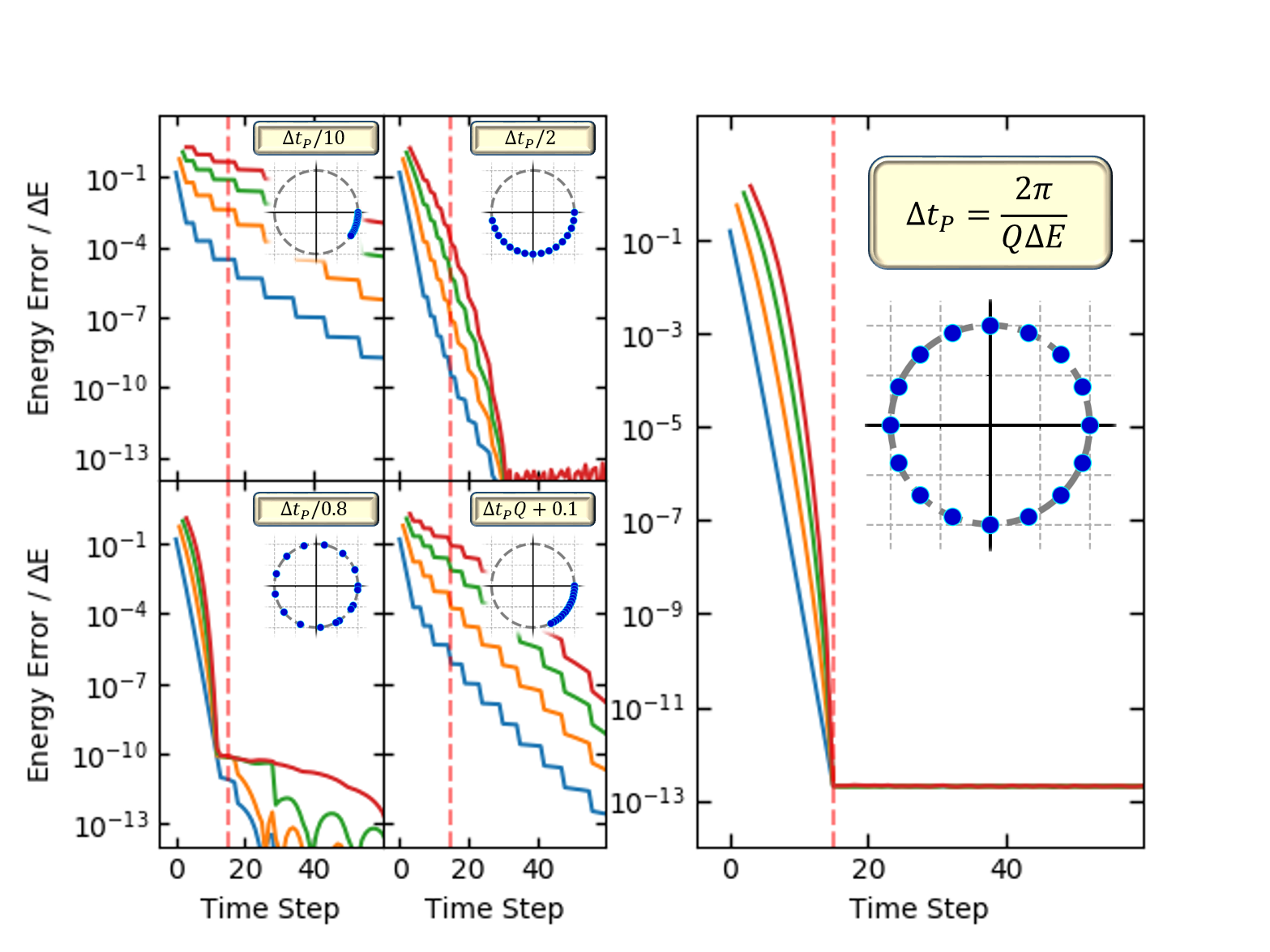}
    \caption{Relative error of the first four eigenvalues from the real time NOVQE secular equation for a Hamiltonian of linear spectrum $E_N = N\Delta E$ ($\Delta E = 0.75$ here) and different time steps $\Delta t$ as a funciton of the number of time steps. The reference state follows $\ket{\Psi_0} \propto \sum_N e^{-E_N}\ket{N}$, such that only the 16 Hamiltonian eigenstates of lowest energy are part of the support space, choosing a coefficient threshold of $10^{-12}$. When solving the secular equation, we choose the same threshold for the SVD decomposition of the overlap matrix. The vertical dashed line marks the 15-th time step, after which we have as many expansion states as vectors in the support space. The large subplot corresponds to the smallest time step which fulfills the phase cancellation condition Eq.~\eqref{eq:PCC}, which reduces to a single condition for this Hamiltonian. The insets in each subfigure correspond to a geometric representation of the phase cancellation condition, with each phase $e^{-it_j\Delta E}$ a point in the unit circle on the complex plane. See text for details.}
    \label{fig:HarmTimeStep}
\end{figure*}

Further, in the limit where long-time evolutions are used the phase cancellation conditions will also be approximately satisfied with high probability as $N_T$ tends to infinity.  To see this, let $t_j = (j+1)/\omega$ where  $\omega^{-1}$ is a uniform random variable on $[0,1/\min(E_N - E_M)]:=[0,\Delta E_{\text{min}}^{-1}]$ (where $M\ne N$).  First we have that
\begin{align}
    &\left|\mathbb{E}\sum_{j=1}^{N_T+1} e^{-i j(E_N-E_M)/\omega} \right|\nonumber\\
    &= \left|\sum_{j=0}^{N_T} \Delta E_{\text{min}}\int_0^{\Delta E_{\text{min}}^{-1}} e^{-i j\omega^{-1}(E_N-E_M)} \mathrm{d}\omega^{-1}\right| \nonumber\\
    & = \left|\sum_{j=1}^{N_T+1} \frac{\Delta E_{\text{min}} (1 - e^{-i j \Delta E_{\text{min}}^{-1} (E_N - E_M)})}{ij(E_N - E_M)}\right|\nonumber\\
    & \in O\left(  \frac{\Delta E_{\text{min}}}{|E_N-E_M|} \sum_{j=1}^{N_T} \frac{1}{j} \right) \nonumber\\ 
    &\subseteq O\left(\frac{\Delta E_{\text{min}} \log(N_T)}{|E_N-E_M|}\right).
\end{align}
Intuitively it is reasonable to expect that if the mean is small then with high probability the PCCs should hold approximately.  In order to demonstrate such a concentration for the oscillating functions that we use here we, however, need to also bound the variance.
\begin{align}
&\left|\mathbb{V}\sum_{j=1}^{N_T+1} e^{-i j(E_N-E_M)/\omega} \right| \nonumber\\
&\le \mathbb{E}\left(\sum_{j\ne k}  e^{-i(j-k) (E_N-E_M)\omega^{-1} t}\right)+(N_T+1)\nonumber\\
&\in O\left(N_T + \frac{\Delta E_{\text{min}}}{|E_N-E_M|}\sum_{j\ne k} \frac{1}{|j-k|} \right) \nonumber \\ &\in O\left(\frac{\Delta E_{\text{min}} N_T \log(N_T)}{|E_N-E_M|} \right).
\end{align}
Thus from Chebyshev's inequality we have that with high probability $\sum_{j} e^{-i j (E_N-E_M)\omega^{-1}}$ will be within
\begin{equation}
     O\left(\sqrt{\frac{N_T\log(N_T)\Delta E_{\text{min}}} {|E_N-E_M|}} \right),
\end{equation}
of the expectation value.
Thus the phase cancellation condition's error for the $N,M$ component is in
\begin{equation}
    O\left(\sqrt{\frac{\log(N_T)\Delta E_{\text{min}}}{N_T|E_N-E_M|} }\right).
\end{equation}
Thus the value of $N_T$ needed to ensure that the PCC holds within error at most $\epsilon$ (with high probability) obeys
\begin{equation}
    N_T \in \widetilde{O}\left(\frac{\Delta E_{\text{min}}}{|E_N -E_M|\epsilon^2} \right).
\end{equation}
Here $\widetilde{O}(\cdot)$ denotes an asymptotic upper bound with multiplicative polylogarithmic factors neglected.
Thus an approximate solution to the phase cancellation conditions will generically hold for a gapped system.

We exemplify the previous theory on the example of a Hamiltonian of linear spectrum $E_N = N \Delta E$, akin to a harmonic oscillator, in Fig.~\ref{fig:HarmTimeStep}. In this case, the PCCs in Eq.~\eqref{eq:PCC} can be fulfilled exactly by a linear time grid $t_j = j\Delta t_P$ with the perfect time step size $\Delta t_P$ defined as

\begin{equation}
    \Delta t_P = \frac{2\pi}{(N_T + 1)\Delta E}.
    \label{eq:HarmSpecPerfDt}
\end{equation}
Indeed, it is easy to check that in the case of a linear spectrum, a linear time grid with time step size given by Eq.~\eqref{eq:HarmSpecPerfDt} fulfills the PCCs exactly after $N_T = Q - 1$ time steps. This can be accomplished with a single time step size since in the case of a linear spectrum the PCCs effectively reduce to a single condition. This can be seen in the rightmost panel of Fig.~\ref{fig:HarmTimeStep}, where exactly after 15 time steps the first four eigenvalues of the secular equation match the exact eigenvalues to the maximal precision. 
This precision is determined by the singular value threshold, $s_{SV}$, introduced into the general eigenvalue problem, which truncates the singular values of the overlap matrix.
In Fig.~\ref{fig:HarmTimeStep} this precision corresponds to $s_{SV}=10^{-12}$, i.e. midway between double and single machine precision. 
This threshold also determines the support space size $Q$. The reference state in all examples in Fig.~\ref{fig:HarmTimeStep} is defined as $\ket{\Psi_0}\propto\sum_N e^{-E_N}\ket{N}$, excited states being exponentially suppressed. The support space is then defined as the Hamiltonian eigenstates with squared coefficients in $\ket{\Psi_0}$ above $s_{SV}$. The inset in this panel represents the PCCs graphically, as the phases of all eigenstates in the support space perfectly span the unit circle, thus cancelling each other.

The smaller panels on the left of Fig.~\ref{fig:HarmTimeStep} show the outcome of choosing a time step size differing from $\Delta t_p$. Small time step sizes are shown in the upper two subfigures, and would be the natural choice from the Krylov interpretation of VQPE~\cite{Stair2020}. These clearly show a significantly slower convergence than the perfect time step derived from the PCCs, which is easy to explain from the phase distribution on the unit circle in the insets. Only once we cover the unit circle close to homogeneously, thus approximately fulfilling the PCCs, can we extract all eigenstates essentially exactly after the minimal number of time steps (see lower left panel in Fig.~\ref{fig:HarmTimeStep}). Due to the periodic nature of the complex phases in Eq.~\eqref{eq:PCC}, large time steps can result in as poor approximations as short ones, as shown in the lower right panel in Fig.~\ref{fig:HarmTimeStep}. The worst such longer time step sizes correspond to particular integer multiples of $\Delta t_P$, namely $z Q\Delta t_P$, where $z$ is an integer. For these time step sizes, the PCCs in Eq.~\eqref{eq:PCC} cannot be fulfilled, even approximately. These are very particular time steps, and thus, for the linear spectrum, a randomly chosen time step $\Delta t \ge \Delta t_p$ is still likely to provide good results for the Hamiltonian eigenstates.

For general spectra, a single time step size $\Delta t$ in a linear grid is unlikely to fulfill all $Q(Q-1)/2$ PCCs exactly. 
From the above analysis, a valid strategy would be to choose a time step size $\Delta t$ and number of timesteps $N_T$ such that we sample a full period of the slowest oscillation in the support space. This is given, e.g., by the minimal energy gap $\Delta E_{\text{min}}$ if we are interested in all excited states contained in the support space, and the ground state gap $\Delta E_{1,2} = E_2 - E_1$ if we only need an estimate of the ground state.
However, in practical implementations it is advantageous, and sometimes necessary, to limit the total simulation time in order to minimize the error.
At the same time, we have to choose a time step size large enough such that each new state is linearly independent from the previous ones. 
Otherwise no new information is added and the variational Ansatz is not improved (see Fig.~\ref{fig:HarmTimeStep}, upper left hand panel, where the energy decreases in a step-like fashion). 
Reconciling these two notions, we propose the following systematic approach: 
\begin{enumerate}
    \item Choose a small enough time step size such that the energy convergence is step-like. Step-like convergence refers to the situation in which adding a new expansion state, i.e. propagating for an additional time step, does not improve the variational Ansatz, resulting in the same (or slightly worse) energy estimates as before including the new step (c.f. upper left panel in Fig.~\ref{fig:HarmTimeStep}). This happens when the inclusion of the new expansion state produces an overlap matrix which has no additional singular value over the threshold $s_{SV}$. 
    \item Perform the VQPE algorithm using the previously identified small time step size, until the first expansion state resulting in an improvement of the energy estimates is produced.
    \item Plotting the lowest eigenvalue of the previous VQPE simulation as a function of the propagation time will result in a nearly horizontal line. This plateau ends after the addition of the final expansion state, which does improve the energies. The length of this plateau defines a new, larger time step size, which can be used in a new VQPE simulation. 
    \item Repeat simulation with the new time step size. If there is still step-like convergence, go back to 3. Otherwise use this as your simulation time step.
\end{enumerate}
This procedure underlies the fact that for devising practical implementations of VQPE, the guiding principle should be to generate linearly independent expansion states rather than to exactly fulfill the PCCs, which nonetheless are a useful perspective for the theoretical analysis of the algorithm. 
This is the strategy we adopt in the results section below.

\subsection{Towards an Optimal Implementation - Toeplitz Structure of $S$}
\label{ss:toeplitz}

Besides being natural to implement on quantum hardware, and presenting the interesting phase cancellation structure described above, VQPE approaches show a further theoretical advantage: when using a linear time grid $t_j = j \Delta t$, the Hamiltonian and overlap matrices in Eq.~\eqref{eq:HandS} have a restrictive structure, which formally reduces the number of measurements that should be needed to solve the generalized eigenvalue problem. Indeed, as pointed out by Parrish \emph{et. al.} in Ref.~\cite{Parrish2019b}, using the real time expansion set these matrices become Toeplitz, meaning that e.g. $S_{j,k} = S_{j+1,k+1}$. In particular, the concrete expressions read

\begin{equation}
	\begin{split}
		H_{j,k} &= \braket{\Phi_{j,0}|H|\Phi_{k,0}} = \braket{\Psi_0|He^{-iH\Delta t(k-j)}|\Psi_0},\\
		S_{j,k} &= \braket{\Phi_{j,0}|\Phi_{k,0}} = \braket{\Psi_0|e^{-iH\Delta t(k-j)}|\Psi_0},
	\end{split}
	\label{eq:HandS_tevol}
\end{equation}
where we have only used the fact that a time-independent Hamiltonian commutes with itself at all times, relying thus exclusively on time translational symmetry. From Eq.~\eqref{eq:HandS_tevol}, it follows that we can reconstruct the 2 $(N_T+1)\times(N_T+1)$ matrices by measuring only $2(N_T+1)$ overlaps in total. 

Unfortunately, as pointed out in Ref.~\cite{Parrish2019b}, the Toeplitz property of the Hamiltonian matrix is lost in actual quantum hardware implementations, if the time evolution operator is Trotterized. In those cases, the commutativity of $U(t_j)$ with $H$ is lost, and thus we either need to evaluate all $(N_T+1)^2$ Hamiltonian matrix elements separately or transition to a higher-order Trotter formula that better approximates the commutation relations. Nevertheless, the same is not the case for the overlap matrix. As long as the expansion states $\ket{\Phi_{j,0}}$ are constructed using a linear grid with a unitary time evolution approximation

\begin{equation}
    U(\Delta t) = e^{-iH\Delta t}\approx U_a(\Delta t),
    \label{eq:TevolOpApprox}
\end{equation}
the Toeplitz condition will prevail. Here, $U_a(\Delta t)$ is the Trotterized time evolution operator and $\Delta t$ is the common time step size of the linear time grid, such that the approximate expansion states obey

\begin{equation}
    \ket{\Phi^a_{j,0}} =\left( U_a(\Delta t)\right)^j\ket{\Psi_0},
    \label{eq:TevolApprox}
\end{equation}
which in the limit of an exact time evolution operator recovers Eq.~\eqref{eq:PhaseEvol}. Now, if $U_a(\Delta t)$ is unitary, the overlap matrix of the approximated expansion states $\ket{\Phi^a(\Delta t)}$ will clearly be Toeplitz, since 

\begin{equation}
    S^a_{j,k} = \braket{\Phi^a_{0,j}|\Phi^a_{0,k}} = \braket{\Psi_0|\left(U_a(\Delta t)\right)^{k-j}|\Psi_0}.
    \label{eq:ToeplitzWillPrevail}
\end{equation}
Given that gate operations in quantum hardware are naturally unitary, this means that it is always possible to guarantee the Toeplitz condition of the overlap matrix, simply by choosing a linear time grid. This is true of course, for the often invoked first order Trotterization~\cite{Mcardle2020} approximation to the time step evolution $U(\Delta t)$, which is indeed unitary.

\subsection{Towards an Optimal Implementation - Unitary Formulation}
\label{ss:unitary}

It is possible to rewrite this generalized eigenvalue problem in a simpler form, exploiting the particular relationship between the Hamiltonian and overlap matrices in Eq.~\eqref{eq:HandS_tevol}, essentially formulating it equivalently to the classical filter diagonalization problem found in signal processing~\cite{mandelshtam2001fdm}. 
This proves to be the ideal formulation of VQPE for quantum computation.

The main insight relies on substituting the Hamiltonian in the secular equation Eq.~\eqref{eq:SecEq} by the time evolution operator $U(\Delta t) = e^{-i H\Delta t}$. This operator is effectively isospectral with the Hamiltonian, indeed the eigenstates $\ket{N}$ of $H$ fulfill
\begin{equation}
    U(\Delta t) \ket{N} = e^{-i E_N \Delta t} \ket{N}.
    \label{eq:TimeEvolEigStates}
\end{equation}
It is important to note that unlike the Hamiltonian, the time evolution operator is not Hermitian, but unitary, having thus complex eigenvalues of unit modulus. We can therefore write a secular equation for the time evolution operator $U(\Delta t)$ as
\begin{equation}
	\sum_k U(\Delta t)_{j,k} c^I_k = e^{-i\varepsilon_I\Delta t} \sum_k S_{j,k} c^I_k,
	\label{eq:SecEqUnit}
\end{equation}  
where the overlap matrix, eigenvalues $\varepsilon_I$, and expansion coefficients $c^I_j$ are the same as in Eq.~\eqref{eq:SecEq}, and the time evolution matrix elements follow, in the single reference implementation,
\begin{equation}
    U(\Delta t)_{j,k} = \braket{\Phi_{j,0}|U(\Delta t)|\Phi_{k,0}} = \braket{\Psi_0|e^{-i H (\Delta t + t_k - t_j)}|\Psi_0}.
    \label{eq:TevolMatels}
\end{equation}
To transform from eigenvalues of Eq.~\eqref{eq:TimeEvolEigStates} to Eq.~\eqref{eq:SecEq}, $\Delta t$ must also be small enough that we can distinguish a physical $E_N$ value from its unphysical periodic images $E_N \pm 2\pi/\Delta t$.

From Eq.~\eqref{eq:SecEq} to Eq.~\eqref{eq:SecEqUnit}, we have simply reformulated the VQPE problem into an equivalent generalized eigenvalue problem with a unitary matrix. The key simplification for the implementation on quantum hardware relies on the realization that the time evolution matrix elements in Eq.~\eqref{eq:TevolMatels} have the same structure as the overlap matrix elements in Eq.~\eqref{eq:HandS_tevol}. Thus, choosing again the time grid $\{t_j\}$ to be linear, i.e. $t_j = j \Delta t$, the time evolution matrix elements coincide with the overlap matrix elements as
\begin{equation}
    \begin{split}
        U(\Delta t)_{j,k} &= \braket{\Psi_0|e^{-i H \Delta t( 1 + k - j)}|\Psi_0}\\
        &= S_{j,k+1} = S_{j-1,k}.
    \end{split}
    \label{eq:UnitAdvantage}
\end{equation}
The last equality is a manifestation of the Toeplitz structure. Thus, according to Eq.~\eqref{eq:UnitAdvantage}, for linear time grids there is no need to measure the time evolution matrix explicitly, since it can be recovered from the measurements for the overlap matrix plus an additional measurement involving an extra expansion state $\ket{\Phi_{N_T+1,0}}$. In this way, exploiting Eq.~\eqref{eq:UnitAdvantage} and the Toeplitz structure of the overlap matrix, the number of measurements reduces from $2(N_T+1)^2$ to just $N_T+2$. Further, as shown in the previous subsection, the Toeplitz structure prevails when implementing the time evolution operators with a unitary approximation, such as first order Trotterization, making the reduction in number of measurements applicable for real implementation on quantum hardware. 

Intuitively, the unitary formulation of VQPE is the quantum algorithm equivalent to measuring the autocorrelation function $g(t) = \braket{\Psi_0|e^{-iHt}|\Psi_0}$ and analyzing its Fourier spectrum. The overlap matrix elements are essentially sampling $g(t)$ at different points, and one can approximate the underlying spectrum once enough samples are obtained. Thus, we are fundamentally expressing the VQPE algorithm in its most natural language, that of autocorrelation functions. In this work, we implement both the traditional and unitary formulations of the VQPE secular equation.

\subsection{Diminishing the Effect of Noise through Singular Value Decomposition}
\label{ss:noise}

In the previous subsections, we have briefly reviewed the theoretical formalism of VQPE approaches, in the new light of the phase cancellation interpretation, but without considering the effects of noise. We turn our attention now to how the presence of noise, comprising both finite numerical precision on the classical computer and measurement uncertainty from the quantum hardware, limits the final accuracy of the VQPE results. We consider systematic noise, due to \emph{a priori} uncontrollable or unavoidable sources, plus any remaining statistical uncertainty after repeated measurements. To this end, the phase cancellation formalism will simplify the analysis. We will restrict ourselves for simplicity to the single reference implementation, but generalizing our conclusions to the multi-reference case is straightforward.

As shown in Eq.~\eqref{eq:Tevol}, VQPE generates a series of $N_T$ states from a reference $\ket{\Psi_0}$ defined by a time grid $\{t_j\}$. 
In a noiseless simulation, any given time grid is more likely than not to produce a set of $N_T$ linearly independent vectors. For them to be linearly dependent requires the following determinant to vanish exactly

\begin{equation}
	\begin{vmatrix}c_0  & c_0 e^{-i E_0 t_1} & c_0 e^{-i E_0 t_2} & \cdots \\
                            c_1  & c_1 e^{-i E_1 t_1} & c_1 e^{-i E_1 t_2} & \cdots \\ 
	                       c_2  & c_2 e^{-i E_2 t_1} & c_2 e^{-i E_2 t_2} & \cdots \\ 
	                       \vdots & \vdots & \vdots & \vdots \\
                             c_{N_T}  & c_{N_T} e^{-i E_{N_T} t_1} & c_{N_T} e^{-i E_{N_T} t_2} & \cdots \\\end{vmatrix} = 0.
	\label{eq:LinDepCondition}
\end{equation}
This equation is one constraint on $N_T$ unknowns $\{t_j\}$, which is generically satisfied by an $(N_T - 1)$-dimensional manifold of $\{t_j\}$ embedded in $\mathbb{R}^{N_T}$. For a linear grid $t_j = j \Delta t$, the choice of time step size $\Delta t$ will generically cause linear dependencies on a subset of $\mathbb{R}$ with measure zero. For example, $\Delta t = \frac{2\pi n}{E_2 - E_1}$ causes a linear dependency in the case of $N_T = 2$. Thus, with the exception of Hamiltonians with a restricted spectrum such as $E_j = j \Delta E$, it seems safe to assume that in almost any time grid chosen, a noiseless simulation will generate $N_T$ linearly independent vectors. Since all the expansion states share the same support space~\footnote{After all, time evolution does not change the absolute values of the expansion coefficients of $\ket{\Psi_0}$ in the Hamiltonian eigenbasis}, a noiseless simulation with $Q$ steps, $Q$ being the size of the support space, should recover all eigenstates exactly. We will assume a linear time grid with time step size $\Delta t$ henceforth.

This ideal notion stops holding the moment we consider noise, both from numerical and measurement origins. 
Noise can for example make states close to linearly dependent, and thus introduce errors in the eigenvalues $\varepsilon_I$ of the secular equation. We will quantify noise introducing the parameter $\epsilon$. Two measured or computed values $\alpha,\beta$ are only distinguishable if $|\alpha - \beta| > \epsilon$. Noise becomes important, for example, in the small time step size limit. When $\Delta t \Delta E_{\text{min}}$ is small, where $\Delta E_{\text{min}}$ is the minimal spectral gap in the support space, the first expansion steps will produce states that are only marginally different to the reference $\ket{\Psi_0}$. These will not improve the variational Ansatz if 

\begin{equation}
	\frac{\Delta T_\epsilon}{\hbar} \Delta E_{\text{min}} < \epsilon,
	\label{eq:LinDepCond}
\end{equation}
where we have recovered Planck's constant to make the units clear. If Eq.~\eqref{eq:LinDepCond} is fulfilled, the magnitude of the difference between the expansion state and the reference will fall below the noise threshold, making the new expansion state $U(\Delta T_\epsilon)\ket{\Psi_0}$ useless from a variational perspective. This is the reason behind the step-like decreasing behavior in the small time step panels of Fig.~\ref{fig:HarmTimeStep}. A finite $\epsilon$ thus determines a minimal time step ${\Delta T_\epsilon > \frac{\hbar \epsilon}{\Delta E_{\text{min}}}}$. 

There will be cases where it is hard to generate precise expansion states with the minimal time step size $\Delta T_\epsilon$ required to offset a given noise level. It becomes thus important to prune the Hamiltonian and overlap matrices of numerical and measurement noise. This can be done by means of a singular value decomposition (SVD) of the overlap matrix: Neglecting all singular values bellow some threshold, which should be larger than the magnitude of the noise.  
In the case of measurement error, this noise scales as $1/\sqrt{M}$ where $M$ is the number of samples.
In the majority of this work, we have thus conservatively chosen a threshold of $10^{-1}$, corresponding to $>100$ samples. 
We used such a truncation already in the results shown in Fig.~\ref{fig:HarmTimeStep}. 
This singular value truncation produces effectively a new but smaller expansion basis of $N_{SVD}$ elements. As a consequence, the number $N_T$ is not the significant measure of how much information is collected in the expansion set, and instead $N_{SVD} \le N_T$ becomes the measure to follow. Only when $N_{SVD} = Q$ will the secular equations recover the exact support space spectrum. 

\begin{figure}
    \centering
    \includegraphics[width=0.5\textwidth]{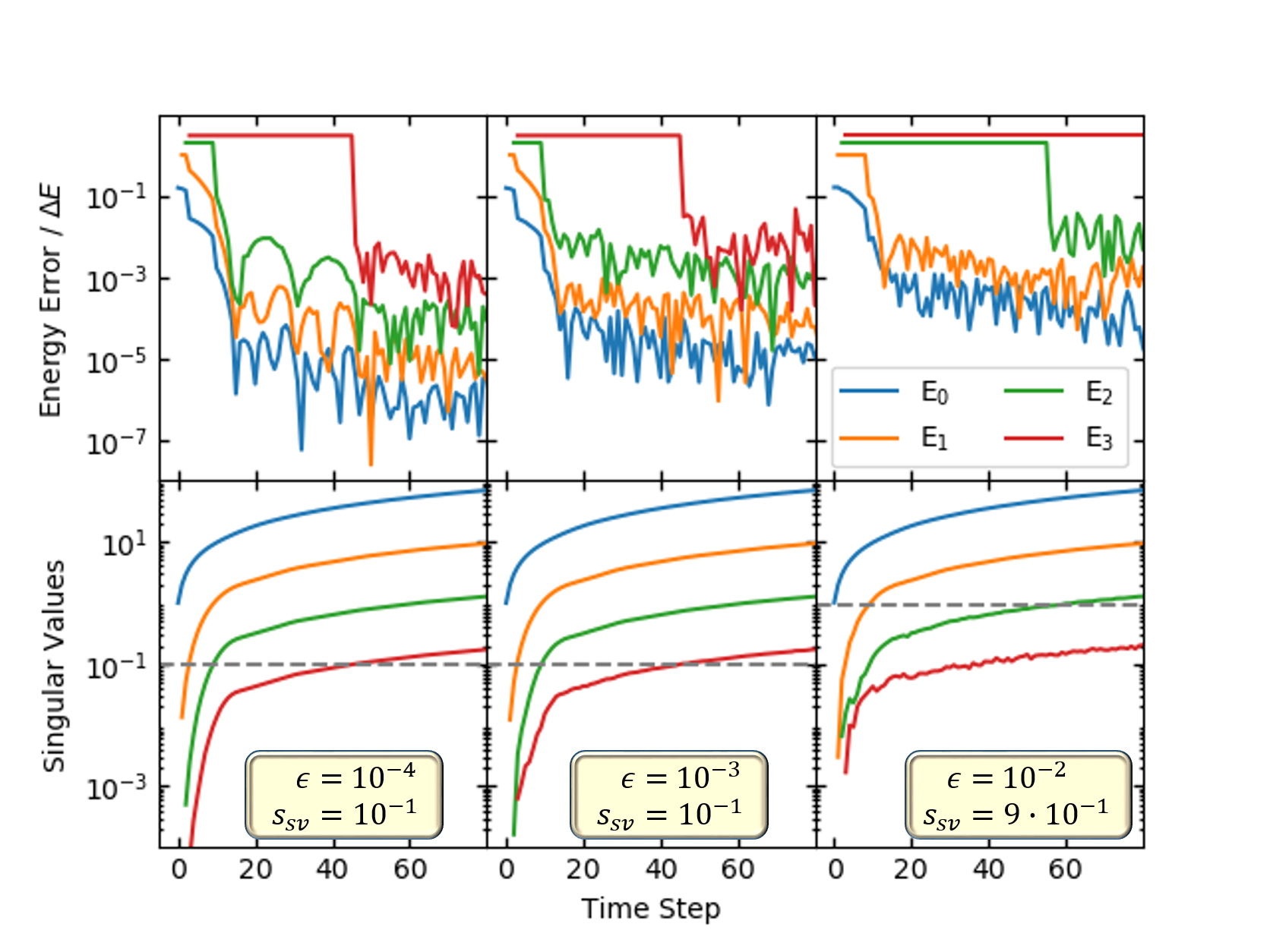}
    \caption{Relative error and corresponding singular values of the overlap matrix for Hamiltonian of linear spectrum with different noise values. Upper panels: Relative error of the first four eigenvalues from the VQPE secular equation for a Hamiltonian of linear spectrum $E_N = N\Delta E$ ($\Delta E = 0.75$ here) and perfect time step $\Delta t_P$, including Gaussian noise $\mathcal{N}(0,\epsilon)$ to the Hamiltonian and overlap matrix elements. We choose the singular value truncation threshold $s_{SV}$ to be at least two orders of magnitude larger than the noise standard deviation~$\epsilon$. The reference state follows $\ket{\Psi_0} \propto \sum_N e^{-E_N}\ket{N}$, the effective support space size determined using the singular value truncation threshold $N^{max}_{SVD} = -\frac{1}{2\Delta E}\ln(s_{SV})-1$. 
    Lower panels: Corresponding singular values of the overlap matrix as a function of time. The horizontal dashed line represents the threshold $s_{SV}$. Note that, until a given singular value is larger than $s_{SV}$, we cannot extract the corresponding eigenvalue from the generalized eigenvalue problem. This is represented by the horizontal lines in the upper panels. See text for details.}
    \label{fig:HarmWithNoise}
\end{figure}

We exemplify this on the harmonic spectrum in Fig.~\ref{fig:HarmWithNoise}, where in the upper panels we show the relative noise error for the first four eigenstates as a function of number of expansion states, introducing Gaussian noise $\mathcal{N}(0,\epsilon)$ of standard deviation $\epsilon$ on the Hamiltonian and overlap matrix elements. We choose the singular value truncation threshold $s_{SV}$ to be between $10^{-1}$ and $1$, and in the lower panels we plot the singular values of $S_{j,k}$ as a function of the number of expansion states, marking $s_{SV}$ as a dashed line. In each simulation, we choose as time step size $\Delta t$ the optimal time step in Eq.~\eqref{eq:HarmSpecPerfDt}, considering a possible support space of 16 elements, regardless of $s_{SV}$~\footnote{One could consider changing the perfect time step size for this harmonic Hamiltonian according to the singular value truncation threshold $s_{SV}$, effectively changing the size of the support space. This would preclude, however, resolving the eigenstate energies to better accuracy than $s_{SV}$, which as shown in Fig.~\ref{fig:HarmWithNoise} is not a true lower bound for the possible accuracy, even in the presence of noise.}. 
In a noiseless simulation, this choice of time step would result in an optimally compact number of effective expansion vectors $N_{SVD}$, which equals the number of actual expansion vectors until the maximal number $N^{max}_{SVD} = Q$ is reached, after which all eigenstates would be resolved accurately. 
The presence of statistical noise has two consequences: on the one hand, the asymptotic accuracy decreases with increasing noise variance. 
As mentioned above, this type of statistical noise can be reduced by sampling.

\begin{figure}
    \centering
    \includegraphics[width=0.5\textwidth]{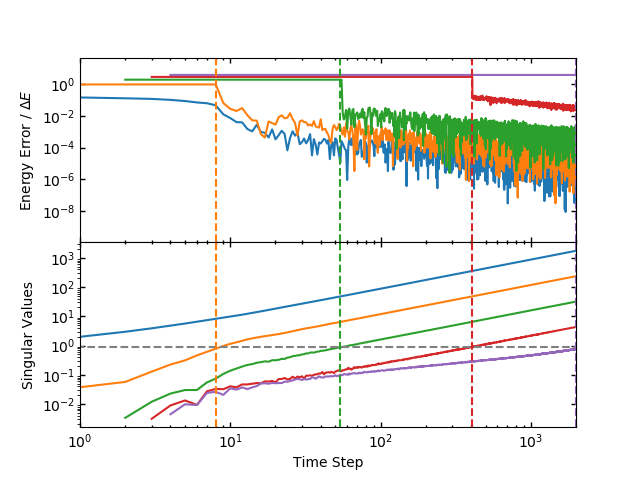}
    \caption{Relative error and corresponding singular values of the overlap matrix for Hamiltonian of linear spectrum for a large enough number of time steps to extract eigenstates below the error threshold. Upper Panel: Relative error of the first five eigenstates from the VQPE secular equation for a Hamiltonian of linear spectrum $E_N = N\Delta E$ ($\Delta E = 0.75$ here) and perfect time step $\Delta t_P$, including Gaussian noise $\mathcal{N}(0,10^{-2})$ to the Hamiltonian and overlap matrix elements. We choose the singular value truncation threshold $s_{SV} = 9\cdot10^{-1}$. The reference state follows $\ket{\Psi_0} \propto \sum_N e^{-E_N}\ket{N}$. Lower Panel: Corresponding singular values of the overlap matrix as a function of time. The horizontal dashed line represents the singular value threshold $s_{SV}$. The time step at which each singular value becomes larger than $s_{SV}$ is marked with a vertical dashed line, connecting upper and lower panels. See text for details.}
    \label{fig:HarmMoreTSteps}
\end{figure}

As a second effect of statistical noise, not all $Q$ eigenstates in the support space are resolved after exactly $Q$ steps, since the corresponding singular values of the overlap matrix fall bellow the truncation threshold $s_{SV}$ (see lower panels of Fig.~\ref{fig:HarmWithNoise}). As long as a given singular value falls bellow $s_{SV}$, the corresponding eigenvalue cannot be extracted from the generalized eigenvalue equation, which is represented in the upper panels of Fig.~\ref{fig:HarmWithNoise} by horizontal straight lines. 
Thus the noise limits what states can be extracted from the reference state $\ket{\Psi_0}$, by setting the minimal singular value truncation threshold $s_{SV}$. Those Hamiltonian eigenstates with smaller absolute coefficient squared than $s_{SV}$ cannot be resolved. However, examining Eq.~\eqref{eq:SopPCC}, we observe that the singular values of the overlap matrix are enhanced linearly with increasing number of expansion states $N_T$, i.e. with increased number of time steps in the VQPE approach, once the PCCs are reasonably fulfilled. 
Thus, it should be possible to extract eigenstates with reference state components below the error threshold by increasing the number of time steps. 
We exemplify this in Fig.~\ref{fig:HarmMoreTSteps}, again on the harmonic spectrum example with exponentially suppressed initial state. 
Because of this choice of starting state, it takes an exponentially large number of extra time steps to resolve every new eigenstate, but it is in principle possible. 
The ideal strategy is of course to propose a reference state with large overlap with the eigenstate of interest, but this discussion shows that it is possible to extract states beyond the dominant one accurately.
Once enough time steps have been produced, any singular value of the overlap matrix can be made to increase above $s_{SV}$, the horizontal dashed line in the lower panel of the figure. 
Of course, this is not unexpected, since the initial starting signal on each excited state is exponentially small by construction.  When a uniform overlap is used as a starting state, then all the SVD values are the same.

We want to address another notion that has been brought up with regards to the effect of noise in VQPE simulations: the overlap matrix condition number $n(S)$~\cite{Stair2020}. 
As shown in~\cite{Stair2020}, typical eigenvalue problems arising in VQPE have extremely large condition numbers, which may suggest high sensitivity to noise.  
This is expected for Hartree-Fock starting states, which ideally have exponentially small overlap with most of the Hilbert space, leading to very large condition numbers.
Still, this can be dealt with by performing an SVD of the overlap matrix and truncating the singular values below $s_{SV}$.
For our application, this reduces the number of linearly independent states we can resolve within our error threshold, thus reducing the best possible accuracy of the results.  
This can be remedied be improved by including additional time evolved states, as discussed above.

In the Supplementary Information, we investigate the role of $n(S)$ for the eigenvalue accuracy in the VQPE approach.
Our results show that, upon pruning the expansion space from the singular values of the overlap matrix below the threshold $s_{SV}$, we consistently obtain accurate eigenvalues in the presence of noise even with matrices of large condition number. 
This is in itself not surprising, since the singular value truncation is effectively proposing an auxiliary generalized eigenvalue problem with smaller condition number.

At this point, we can return our attention to the notion of multi-reference implementations of the VQPE, such as the multi-reference quantum Krylov method of Stair~\emph{et.~al.}~\cite{Stair2020}. In this work the use of several reference states is proposed in order to reduce the condition number of the Hamiltonian and overlap matrices in the expansion space, at the cost of requiring a larger  number of expansion states for the same ground state accuracy. Using the phase cancellation picture, we argue that the worsened ground state energy convergence is due to two distinct, cooperating factors: the increased size of the total support space, and the smaller number of expansion vectors in each individual support space. In the multi-reference formulation of VQPE, each state $\ket{\Psi_I}$ has its own support space w.r.t. $H$, which we refer to as individual support spaces, the union of these forming the total support space of the implementation. The individual support spaces will be in general distinct from each other. Clearly, the larger total support space allows for a more flexible variational ansatz, from which it is possible to extract more Hamiltonian eigenstates than in the single reference case. However, this comes at the price of requiring more expansion states to perform the phase cancellation procedure to purify individual eigenstates. From our results, performed in classical simulations with noise, and on actual noisy quantum hardware, the larger condition numbers do not result in large errors in the eigenvalue estimates $\varepsilon_I$, and thus we conclude the condition number alone should not be a reason to employ multi-reference VQPE implementations. However, in some  excited states simulations a multi-reference approach might accelerate covergence, in the same way that band Lanczos improves normal Lanczos in this regard~\cite{Koch2011,Meyer1989}.

The previous considerations hold for statistical errors but a more careful analysis needs to be performed for systematic errors in the implementation of the Hamiltonian dynamics. 
For example, in the case of a Trotterized Hamiltonian, the reference states are created under the evolution of a Hamiltonian different than the one of interest, which will result in errors in the eigenvalue estimation which cannot be reduced through sampling.

\subsection{Comparison between unitary VQPE and QPE}
\label{ss:vqpeVSqpe}

Here we compare the unitary formulation of VQPE in Eq.~\eqref{eq:TimeEvolEigStates} to conventional QPE in the general case of a multi-dimensional support space.
For the special case of a 1-dimensional support space, there are adaptive variants of QPE that use one ancilla qubit and achieve Heisenberg-limit measurement~\cite{wiebe2016}.
Recently developed methods~\cite{somma2019quantum,lin2021heisenberg} devise QPE variants that use one ancilla qubit and are suitable for larger support spaces (which is also the case for VQPE).
In an adaptive approach where $| \Psi_0 \rangle$ is the ground state, a different $t$ would be chosen for each measurement to maximize the extraction of information about $E_0$ rather than performing multiple measurements to estimate the expectation value of $g(t) = \langle \Psi_0 | e^{-iHt} | \Psi_0 \rangle = e^{-i E_0 t}$ for a single choice of $t$.

Quantum phase estimation (QPE) is a natural algorithm to compare VQPE against; however, there are a wealth of different phase estimation algorithms known in the literature and further some applications of phase estimation can even be used in concert with VQPE.  Our aim in this section is to compare and constrast different flavors of phase estimation to VQPE and also show how QPE can be used to accelerate learning the expectation values of the VQPE circuit through amplitude estimation.

There are broadly two categories of phase estimation algorithms, iterative phase estimation and Fourier-based phase estimation.  Fourier-based phase estimation is perhaps the best understood approach to performing phase estimation.  An advantage of this approach is that it is known to precisely achieve optimal scaling of the uncertainty with the number of applications of the underlying unitary (i.e. it saturates the Heisenberg limit~\cite{giovannetti2006quantum,giovannetti2011advances}).  The optimal approach to Fourier-based phase estimation deviates slightly from traditional approaches by using an optimized initial state which deviates from the Fourier state typically used in older approaches.  Specifically, the input state is taken to be an $m$-qubit state of the form
\begin{equation}
    \chi\ket{0}= \sqrt{\frac{2}{2^m+1}} \sum_{n=0}^{2^m-1} \sin\left(\frac{\pi(n+1)}{2^m+1} \right)\ket{n}.
\end{equation}
This state is chosen to minimize an estimate of the circular variance, known as the Holevo variance, of the eigenphases of the unitary $e^{-iHt}$ that results from the phase estimation protocol.  Next let us define notation for a controlled directional evolution below

\[
\Qcircuit @C=1em @R=.7em {
& & \gate{\dagger} & \qw  & &&\ctrlo{2}&\ctrl{2}&\qw\\
&    &               &       &=&   &&&&&&\\
& & \gate{U}\qwx[-2]&\qw  & &  &\gate{U}&\gate{U^\dagger}&\qw
}
\]
This notation is useful here because in quantum phase estimation for the simulation of electronic structure, the cost of performing a controlled-directional evolution as seen above is approximately the same as performing an ordinary controlled evolution. 
Applying phase estimation with a circuit of this form provides twice as much phase per application of the unitary as a controlled-$U$ application would experience.  
The uncertainty from this phase estimation circuit, as quantified by the Holevo variance, is approximately $\epsilon_{PE} = \frac{\pi}{2^{m+1}}$, where $m$ is the number of qubits used in the QPE register.  This is precisely the Heisenberg limit. 
\begin{figure}
    \centering
    \[\Qcircuit @C=1em @R=.7em { &\lstick{\ket{0}}&\multigate{3}{\chi}&\gate{\dagger}&\qw&\qw&\qw&\multigate{3}{QFT^{-1}}\\
    &\lstick{\ket{0}}&\ghost{\chi}&\qw&\gate{\dagger}&\qw&\qw&\ghost{{QFT^{-1}}} \\
    &\lstick{\ket{0}}&\ghost{\chi}&\qw&\qw&\gate{\dagger}&\qw &\ghost{{QFT^{-1}}}\\
    &\lstick{\ket{0}}&\ghost{\chi}&\qw&\qw&\qw&\gate{\dagger}&\ghost{{QFT^{-1}}}\\
    &\lstick{\ket{\psi}}&\qw&\gate{e^{-iHt}} \qwx[-4]&\gate{e^{-i2Ht}}\qwx[-3]
    &\gate{e^{-i4Ht}}\qwx[-2]&\gate{e^{-i8Ht}}\qwx[-1]&\qw}
    \]
    \caption{Four qubit version of asymptotically optimal phase estimation circuit.}
    \label{fig:my_label}
\end{figure}
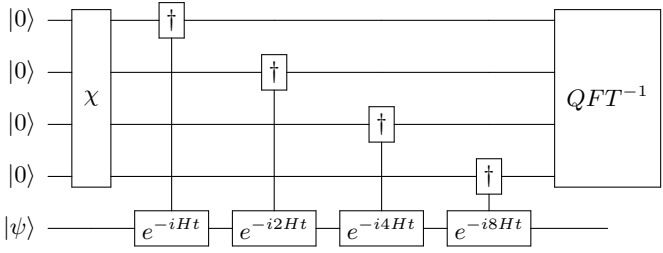

Of course, error in the simulation dynamics needs to be considered in any such simulation.  For simplicity we will consider using the lowest-order Trotter-Suzuki formula within the phase estimation protocol.  In the limit where the uncertainty in the eigenphase is small relative to $\pi$, the Holevo variance approximately corresponds to the variance.  Then since the Trotter-Suzuki error is uncorrelated with the phase estimation error, the total error in the phase estimate that arises from using a timestep of duration $t$ is
\begin{align}
    \epsilon  &\le  \frac{\sqrt{\epsilon_{PE}^2 + \pi^2\epsilon_{TS}^2(t)}}{t}\\
    &\approx \frac{\pi\sqrt{\frac{1}{2^{2m+2}} + \epsilon_{TS}^2(t)}}{t} .
\end{align}
This suggests that if we choose both error sources to be $\epsilon t /(2\sqrt{2}\pi)$ then the overall error will be $\epsilon$.  This corresponds to
\begin{equation}
    m = \lceil-\frac{1}{2} + \log_2\left(2\pi/\epsilon t\right)\rceil \le \log_2\left(2\sqrt{2}\pi/\epsilon t\right) .
\end{equation}

The error in the Trotter-Suzuki formula is difficult to estimate a priori and for the purposes of simplifying the comparison we will choose the symmetric Trotter formula.  Using the results of Proposition 16 from~\cite{childs2019theory} we have that if $H=\sum_j \alpha_j H_j$ then the error in the symmetric Trotter formula is upper bounded by
\begin{align}
    \epsilon_{TS}(t) \le& \frac{t^3}{12} \sum_j \sum_{p>j} \sum_{q>j} \alpha_p \alpha_q \alpha_j \|[H_q,[H_p,H_j]\|\nonumber\\
    &+ \frac{t^3}{24} \sum_j \sum_{p>j} \alpha_p \alpha_j^2 \|[H_j,[H_j,H_p]\|  .
\end{align}
This bound can be further simplified using the fact that each of the Hamiltonian terms $H_j$ in quantum chemistry is of norm $1$.  Therefore let us denote $S$ to be the set of all tuples $(p,q,j)$ such that $[H_q,[H_p,H_j]]\ne 0$.  Then we can further bound
\begin{align}
    \epsilon_{TS}(t) \le& \frac{2t^3}{3} \sum_j \sum_{p>j} \sum_{q>j} |\alpha_p \alpha_q \alpha_j| \delta_{(q,p,j)\in S}\nonumber\\
    &+ \frac{t^3}{3} \sum_j \sum_{p>j} |\alpha_p \alpha_j^2| \delta_{(j,p,j)\in S} \nonumber\\
    &:= \Gamma^3 t^3.\label{eq:gambd}
\end{align}
Under the assumption that the error from QPE is equal to the Trotter error we then find that a sufficient choice for the Trotter-step is
\begin{equation}
    t = \Gamma^{-3/2}\sqrt{\frac{\epsilon}{2\sqrt{2}\pi}}.
\end{equation}
Therefore the value of $m$ needed in the phase estimation step is
\begin{equation}
    m \le \frac{3}{2}\log_2\left(\frac{2\sqrt{2} \pi \Gamma }{ \epsilon} \right).
\end{equation}
Thus if $M$ is the number of terms in the Hamiltonian then the maximum number of operator exponentials that need to be simulated in the circuit to achieve Holevo variance $\epsilon$ is
\begin{equation}
    N_{\exp} \le 2M\left(\frac{2\sqrt{2} \pi \Gamma }{ \epsilon}\right)^{3/2}.
\end{equation}
This process needs to be repeated a number of times to find an eigenstate $j$ which occurs with a probability of $|\psi_j^0|^2$.  This implies that the total number of repetitions needed to reconstruct the entire spectrum with high probability is in $\widetilde{O}(1/\min_j |\psi_j|^2)$ and thus
\begin{equation}
    N_{\exp,QPE} \in \widetilde{O}\left( \frac{M}{\min_j |\psi_j^0|^2}\left(\frac{ \Gamma }{ \epsilon}\right)^{3/2}\right).
\end{equation}
Note that the performance of this algorithm can be improved using optimized simulation methods and the use of amplitude amplification, but we forgo these optimizations here since they are less appropriate for experiments in the NISQ era.

Now let us consider the analogous problem for estimating the corresponding eigenvalue using VQPE.  The first step involves learning the matrices $H_{j,k}$ and $S_{j,k}$.  In order to do so, we need to apply $e^{-iHt_j}$ for all $j$ considered.  There are of course a host of quantum simulation algorithms that can be employed to perform this simulation.  For simplicity, we will consider the second-order Trotter-Suzuki formula.  Consider~\eqref{eq:gambd}.  From this expression we have that we can perform $e^{-iH t_j}$ using $O(M(\Gamma t_{\max})^{3/2}/\sqrt{\epsilon_{TS}})$ operator exponentials, where $|t_j|\le t_{\max}$ for all $t_j$.  Thus we can use this circuit and the Hadamard test to compute the real and imaginary components of $\langle\phi_j|\phi_k\rangle$ within error at most $\epsilon$ with high probability (using the Chernoff bound to justify the high-probability statement) using
\begin{equation}
    N_{exp,S_{ij}} \in \widetilde{O}\left(\frac{M(\Gamma t_{\max})^{3/2}}{\epsilon^{5/2}} \right). 
\end{equation}
Similarly, using the arguments of~\cite{wecker2015progress} we have that the number of operator exponentials needed to approximate the mean energy within the same error requirements is in
\begin{equation}
    N_{exp,H_{ij}} \in \widetilde{O}\left(\frac{M(\sum_j |\alpha_j|)^2(\Gamma t_{\max})^{3/2}}{\epsilon^{5/2}} \right). 
\end{equation}
Next, using the fact that the induced $2$-norm is at most $N_T$ times the max-norm, we have that if $\widetilde{S}$ and $\widetilde{H}$ are the approximated matrices then we need to take $\epsilon \rightarrow \epsilon/N_T$ in order to ensure that the total error in the reconstructed matrix (as measured by the operator norm) is at most $\epsilon$.  This implies that the number of exponentials needed to reconstruct both matrices within the required error budget is in 
\begin{equation}
    N_{exp,S} \in \widetilde{O}\left(\frac{N_T^{9/2}M(\Gamma t_{\max})^{3/2}}{\epsilon^{5/2}} \right), 
\end{equation}
\begin{equation}
    N_{exp,H} \in \widetilde{O}\left(\frac{N_T^{9/2}M(\sum_j |\alpha_j|)^2(\Gamma t_{\max})^{3/2}}{\epsilon^{5/2}} \right). 
\end{equation}
Then from the Hellman-Feynman theorem we have that the eigenvalues of both matrices lie within an $\epsilon$-neighborhood of the original eigenvalues.  Next using standard matrix inequalities for the error in the matrix inverse~\cite{horn} (under the assumption that $\epsilon$ is less than the minimum eigenvalue gap) yields that the error in any eigenvalue reconstructed from~\eqref{eq:SecEq} has error in $O(\|H\| \|S^{-1}\|^2 \epsilon)$. Thus if we wish the error in the reconstructed eigenvalues to be at most $\epsilon$ then we finally need to take $\epsilon\mapsto \epsilon /\|H\| \|S^{-1}\|^2$.  This, combined with the bound that $\|H\| \le \sum_j |\alpha_j|$ leads to a final complexity scaling of the number of exponentials needed to reconstruct the eigenvalues within error at most $\epsilon$ with high probability is in 

\begin{equation}
    N_{exp} \in \widetilde{O}\left(\frac{M(N_T\sum_j |\alpha_j|)^{9/2}\|S^{-1}\|^5(\Gamma t_{\max})^{3/2}}{\epsilon^{5/2}} \right).
\end{equation}
Asymptotically, this analysis suggests that VQPE may be favorable in cases where \begin{equation}
\min_j |\psi_j^0|^2 \ll \frac{\epsilon}{(N_T \sum_j |\alpha_j|)^{9/2} \|S^{-1}\|^5},
\end{equation}
due to the potential benefits that could arise from inferring having to sample low probability events using phase estimation.  However, such asymptotic analysis does not conclusively show which method will be superior as both are upper bounds that are potentially loose.  For this reason, numerical studies such as the one undertaken in this paper, are essential for gauging the actual performance of these algorithms.

For Toeplitz matrices such as in Eqs.~\eqref{eq:HandS_tevol} and \eqref{eq:UnitAdvantage} to be diagonal in a Fourier basis, they must have additional circulant matrix structure satisfying
\begin{equation}
    S_{j,N_T} = S_{j+1,0},
\end{equation}
 which corresponds to an aliasing condition whereby all energy eigenvalues $E_N$
 are an integer multiple of a base energy,
\begin{equation}
    E_N/\omega_1 \in \mathbb{Z}.
\end{equation}
An accurate approximation of this condition requires a very small value of $\Delta t$ and thus a large number of ancilla qubits in the QPE circuit and large $N_T$.
There is a formal equivalence between QPE and unitary VQPE when this condition is exactly satisfied, although this occurs beyond the intended small-$N_T$ operating regime of any VQPE variant.
Practically, VQPE and QPE remain distinct in that VQPE uses measurements of one ancilla qubit to statistically estimate overlap matrix elements, while QPE uses measurements of many ancilla qubits to sample from approximate eigenpairs.

Besides requiring fewer ancilla qubits, the main benefit of VQPE over QPE in the small-$N_T$ regime is its exact diagonalization of a generalized eigenvalue problem rather than relying on approximate diagonalization within a Fourier basis.
The overlap matrix of this eigenvalue problem has a numerical rank corresponding to the size of the support space, and this reduced rank is not reliably exposed by the diagonal elements of the overlap matrix in the Fourier basis.
As a result, VQPE achieves a higher effective energy resolution than QPE for small $N_T$ values and an even smaller support space by exploiting an SVD-based projection of the eigenvalue problem into a numerically relevant subspace.

\begin{figure}
    \centering
    \includegraphics[width=0.5\textwidth]{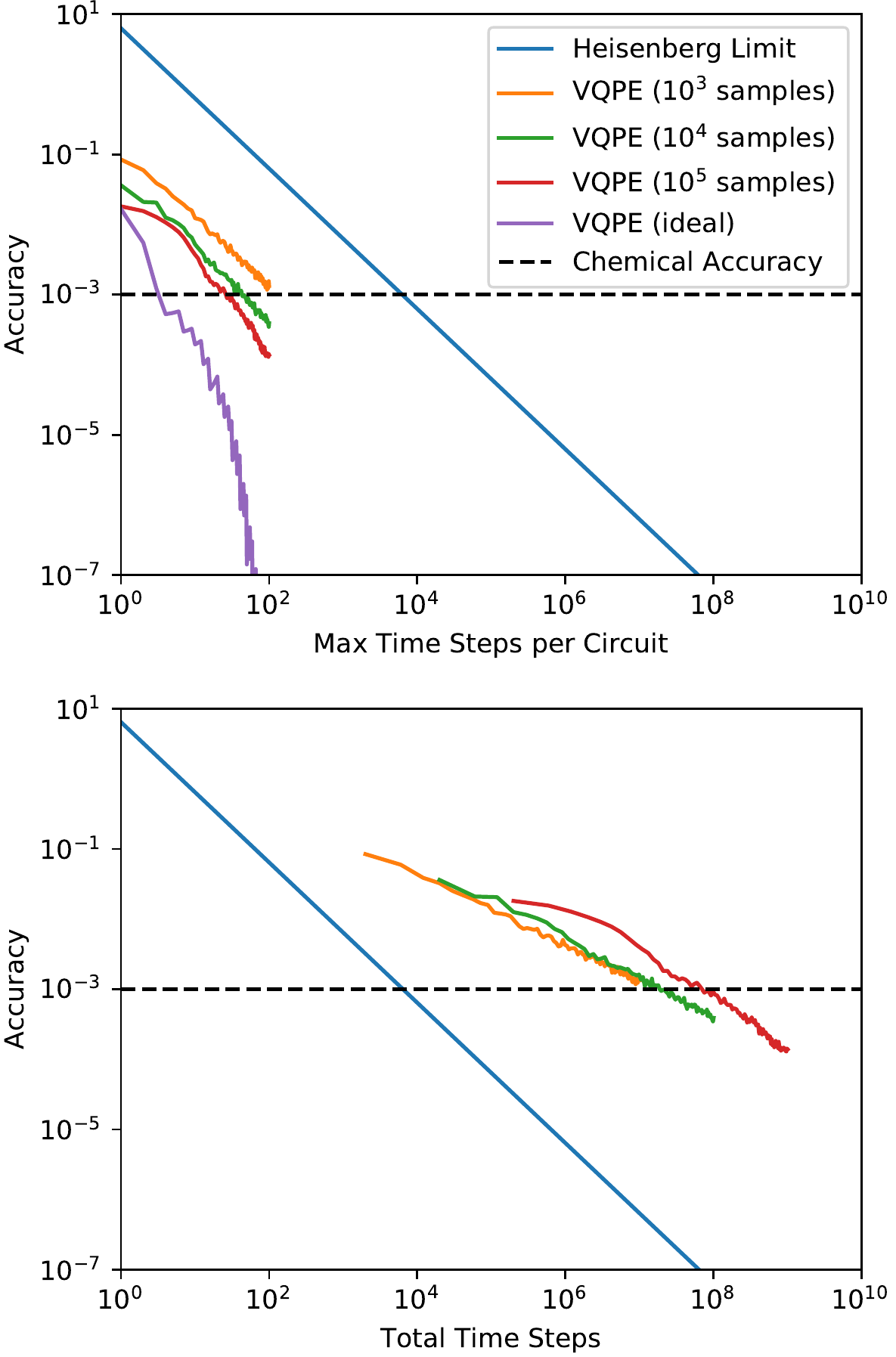}
    \caption{Accuracy comparison between VQPE and the Heisenberg limit for the ground state energy of LiH. Varying numbers of samples per expectation value are shown for VQPE alongside the idealization of VQPE without finite-sampling errors.
    We use SVD cutoffs of 2, 0.9, and 0.3 respectively corresponding to $10^3$, $10^4$, and $10^5$ samples and $10^{-6}$ for the ideal case.
    All methods use $\Delta t = 0.5$. The top and bottom plots show the the maximal evolution time (i.e. how many time steps need to be included for energy convergence), related to the circuit depth, and the total evolution time, i.e. the sum of all time segments needed.} 
    \label{fig:QPE_comparison}
\end{figure}

\begin{figure}
    \centering
    \includegraphics[width=0.5\textwidth]{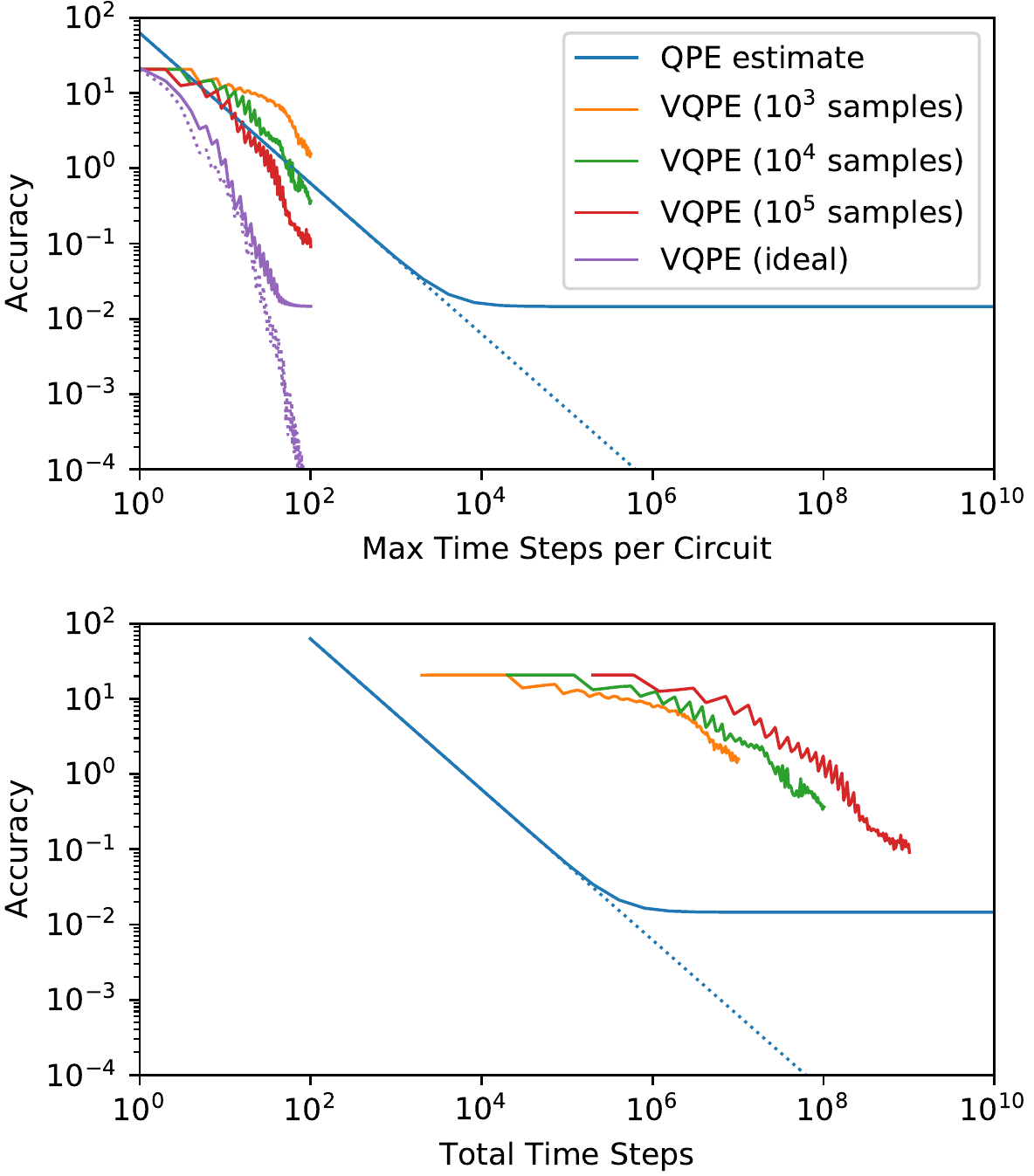}
    \caption{Accuracy comparison between VQPE and QPE for the ground state energy of the 10-qubit transverse field Ising model with first-order Trotterization and $\Delta t = 0.05$.
    We use SVD cutoffs of 1, 0.6, and 0.2 respectively corresponding to $10^3$, $10^4$, and $10^5$ samples and $10^{-5}$ for the ideal case.
    Dotted lines show QPE and idealized VQPE results without Trotter errors.
    The reference state, $|\Phi\rangle = |\phi\rangle^{\otimes 10}$, is a product
    state the maximizes overlap with the exact ground state,
    $|\langle \Psi_0 | \Phi \rangle|^2 \approx 0.014$ for $|\phi\rangle \approx 0.979|0\rangle + 0.205|1\rangle$.} 
    \label{fig:QPE_comparison2}
\end{figure}

In Fig.~\ref{fig:QPE_comparison}, we plot two different resource  (for LiH discussed further in Section~\ref{section:methods}) estimates that exemplify the near-term benefits of VQPE and the long-term benefits of QPE, here shown as the Heisenberg limit.
VQPE is able to achieve substantially higher accuracy than idealized QPE for the same number of time steps per circuit, thus it is better suited for utilizing near-term hardware with shorter circuit depths and fewer qubits available for use as ancillae.
However, QPE utilizes its deeper circuits to achieve Heisenberg-limited energy resolution, which is more efficient in overall run time for achieving high accuracy.  But, as with modern parallel algorithms, it is clear that each matrix element can be calculated independently and thus can be run as a parallel algorithm over multiple quantum computers. 
The expected QPE run time is also amplified by $|\langle \Psi_0 | 0 \rangle|^{-2} \approx 1.02$ to account for the probability of collapsing into a state other than the ground state, which is a negligible overhead in this example.
Other than finite-sampling costs and errors, this example does not consider other sources of error such as Trotterization or quantum circuit noise, which can further degrade the accuracy of both QPE and VQPE.
For example, Fig.~\ref{fig:QPE_comparison2} depicts a $\Delta t$-dependent error floor induced by Trotterization errors (for the transverse field Ising model defined in Eq.~\eqref{eq:tfim} and discussed further in Section~\ref{section:methods}) that remains overwhelmed by sampling errors in VQPE.
With shorter circuit depths and fewer ancilla qubits, VQPE is less susceptible than QPE to these extrinsic sources of error.

\subsection{Inclusion of Other Time-Evolved States}
\label{ss:other}

\begin{figure}[htb]
  \centering
  \includegraphics[width=0.5\textwidth]{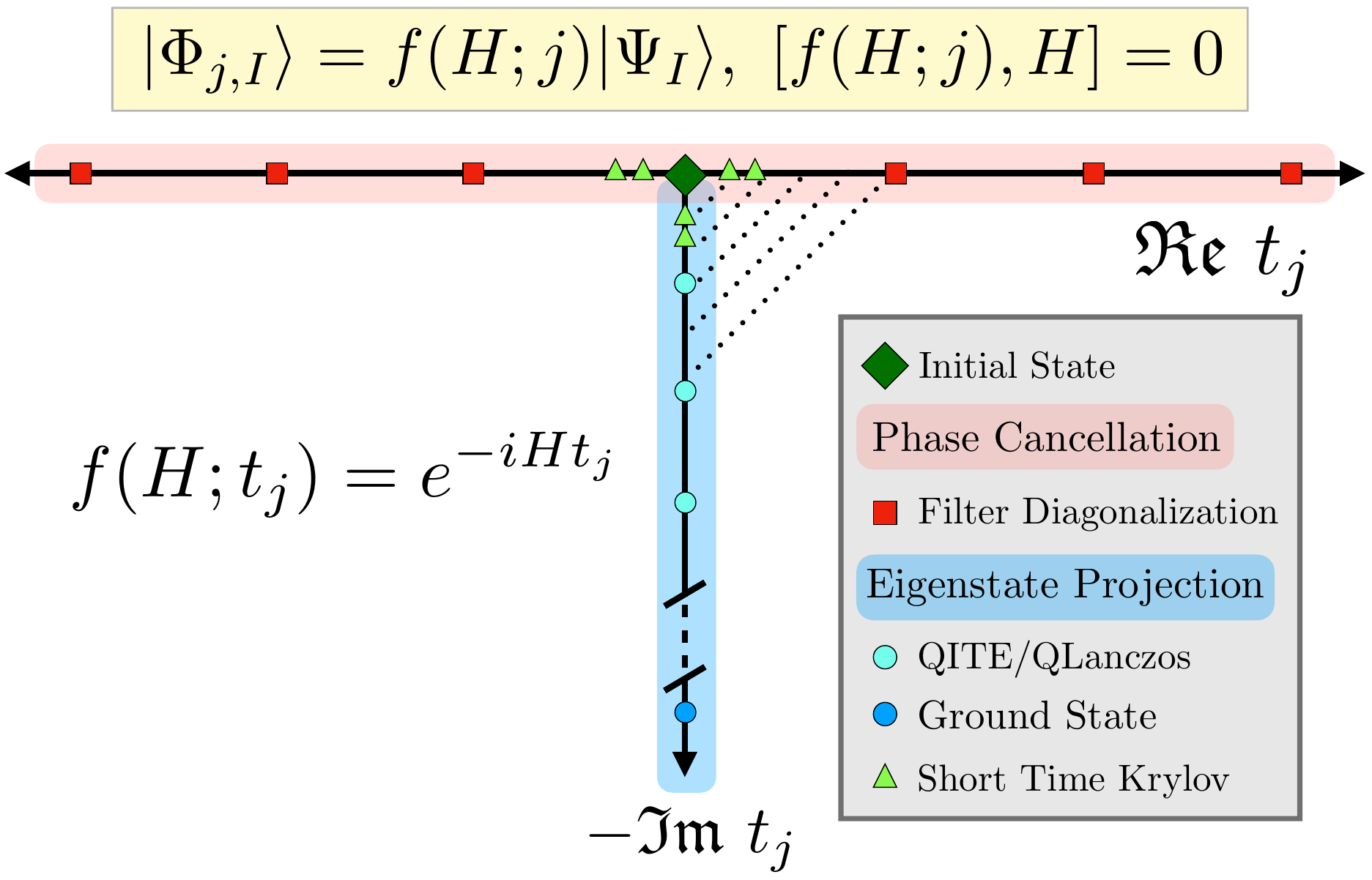}
  \caption{A schematic picture of recent eigenvalue algorithms relying on time evolution proposed for quantum computers. The plane corresponds to the complex time plane, the horizontal axis being real time, the vertical axis imaginary time.  The colored dots denote expansion states $| \phi_{j,I} \rangle$. 
  The operator $f(H;j)$ denotes any unitary that commutes with the Hamiltonian, and thus the expansion states formed by applying this unitary will remain in the support space and can be used to extract eigenstates.
  The real time axis is where the phase cancellation picture applies. On the imaginary time axis, the coefficients of the excited states are suppressed as time increases, leading to the ground state. 
  The dotted lines indicate that short time imaginary and real time states will be very similar, as argued in the short time Krylov method~\cite{stair2020multireference}. 
  We emphasize that combinations of states generated on the entire real/imaginary time plane have essentially the same support space and can be used as a basis to solve the generalized eigenvalue problem.} 
  \label{fig:schematics}
\end{figure}

In this work we focus on real-time evolution. 
However, in practice any time-evolved state shown in Fig~\ref{fig:schematics}, real or imaginary, could be included in the expansion set for solving the generalized eigenvalue equation. 
Moreover, any unitaries $f(H;j)$ that commute with the Hamiltonian would be viable candidates to produce an expansion set with the following important property: all the states in the expansion set share an identical support space w.r.t the Hamiltonian.
Thus any expansion $\ket{\Phi_{j,I}}=f(H;j)\ket{\Psi_I}$ formed by any such unitary will eventually cover the whole support space without introducing more eigenstates.
Recent algorithms such as QLanczos~\cite{Motta2020b,Yeter2020}, quantum filter diagonalization~\cite{Parrish2019b}, quantum Krylov approaches~\cite{Stair2020}, and quantum power methods~\cite{kyriienko2020quantum,seki2021quantum} have all proposed using real or imaginary time evolution in various ways to generate states for creating a wave function Ansatz.  
The real/imaginary time evolution plane is illustrated in Figure~\ref{fig:schematics}, along with the states used in these various algorithms.   
As long as the support space is retained in the time evolution and the states are linearly independent, efficient eigenvalue extraction will be possible. 
There are several benefits derived from using only the real-time axis which include only a linear number of measurements for setting up the generalized eigenvalue problem, almost no optimization of parameters for determining the time evolution, and a well conditioned problem for extracting information about the ground and low lying excited states. 

\section{Methodological Details}
\label{section:methods}

We turn now to presenting concrete numerical data for Hamiltonians from condensed matter physics and quantum chemistry.
For practical applications, the most important quality of the VQPE approach is doubtlessly its compactness. 
By this, we refer to the number of variational parameters required to obtain accurate eigenvalues. 
The number of variational parameters is $(N_T + 1) N_R$, where $N_R$ is the number of reference states (for all examples shown below, we use a single reference $N_R=1$), and we choose chemical accuracy ($\sim 1.6$ mHa) as our target accuracy. 
Thus, we apply VQPE to several many-body systems of different complexity, taken from the fields of quantum chemistry and condensed matter physics: several small molecules, including weakly correlated ones, LiH (3-21g basis set~\cite{binkley1980a}, 11 orbitals and 4 electrons), H$_2$O (cc-pVDZ basis set~\cite{dunning1989a}, 24 orbitals and 10 electrons), N$_2$ (cc-pVDZ basis set~\cite{dunning1989a}, 28 orbitals and 14 electrons), two moderately correlated, linear H$_6$ (STO-6g basis ~\cite{hehre1969a}, bond length 1.5~\AA, 6 orbitals and 6 electrons) and  C$_2$ (cc-pVDZ basis set~\cite{dunning1989a}, 28 orbitals and 12 electrons), and one strongly correlated Cr$_2$ (def2-SVP basis set~\cite{weigend2005a}, 30 orbitals and 24 electrons), and the transverse field Ising model (2 site on hardware, 10 site on simulator). 
We simulate the VQPE calculation for the molecular systems classically, while running the Ising model calculations on IBM's quantum hardware and simulator. Further details, such as molecular geometries, can be found in the Supplementary Material.

For the classical simulations, we perform time evolution using exact diagonalization (ED) dynamics based on the Lanczos algorithm~\cite{Park1986, Manmana2005,Kollath2007}. 
In the case of all molecular systems except for Cr$_2$, the small basis sets allow us to perform the time evolution including all electrons and all orbitals explicitly, while for Cr$_2$ we restrict the simulation to the widely studied 30 orbital active space. 
Still, only for LiH and H$_6$ is it computationally feasible to perform exact dynamics. 
For all other systems, we must truncate the Hilbert space, considering only a subset of all Slater determinants in the corresponding active spaces. 
This presents an approximation, and we perform a finite-size effect study by comparing the dynamics of progressively larger truncations, from one thousand to one million determinants, which we show for Cr$_2$, the most complicated of the systems considered. 
The determinants of each truncation are chosen using the adaptive sampling configuration interaction (ASCI) algorithm~\cite{Tubman2016,Tubman2018a,Tubman2020}. 
This is an iterative selected configuration interaction approach, which explores the Hilbert space and identifies the most important determinants for a ground state approximation, providing highly accurate yet moderately sized truncations. 
Thus, by optimizing the truncated spaces with respect to the ground state description, we can reliably infer the full Hilbert space limit from our finite size calculations. 
In practice, we determined a one million determinant Hilbert space truncation with ASCI for Cr$_2$ ($10^5$ determinants for H$_2$O, C$_2$ and N$_2$), and built all smaller truncations of size $N$ from this one, by picking the $N$ determinants with the largest coefficients in the one million ($10^5$  for H$_2$O, C$_2$ and N$_2$) determinant ground state wave function. 
For all molecular systems, we determined the optimal time step size $\Delta t$ using the strategy described in the theory section (see Fig.~\ref{fig:lih}), and use the Hartree-Fock state as reference state $| \Psi_0 \rangle$. 
For these classical simulations, we did not exploit the unitary formulation of VQPE, instead solving the generalized eigenvalue equation~\ref{eq:SecEq}.

For the simulations on IBM's quantum hardware and simulator, we consider the transverse field Ising Model (TFIM) with open boundary conditions, defined by the following Hamiltonian
\begin{equation}
\label{eq:tfim}
    H = -J \left( \sum_i Z_i Z_{i+1} +h\sum_i X_i \right) \ ,
\end{equation}
where $X, Y$, and $Z$ above and henceforth denote the Pauli operators, $J=1$ corresponds to the spin coupling and $h=2$ the external field. 
In these simulations, we chose a time step size $\Delta t = 0.05$ and approximated the time evolution operator via first order Trotterization~\cite{Mcardle2020}. 
Each run was performed with 8192 shots. 
For all hardware data shown, $| \Psi_0 \rangle$ readout error mitigation was performed.

\begin{figure}
    \centering
    \includegraphics[width=0.3\textwidth]{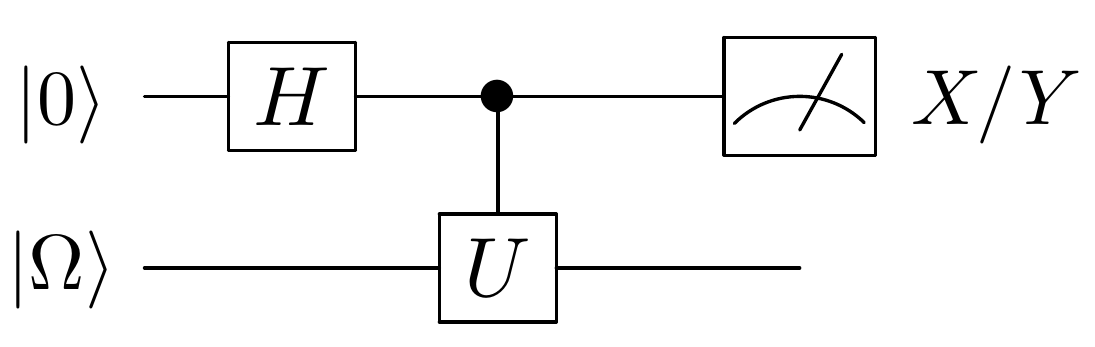}
    \caption{Hadamard test circuit, which computes the real and imaginary parts of the overlap matrix, where $| \Omega \rangle$ denotes the system and the top qubit is the ancilla. Here $U = e^{-itH}$.}
    \label{fig:OverlapCircuit}
\end{figure}

As discussed in the theory section, the Toeplitz property of the overlap matrix means that the number of measurements is linear with the number of timesteps $N_T+2$. Further, we take advantage of the unitary formulation of VQPE to measure only the real and imaginary components of the overlap matrix explicitly, not the Hamiltonian matrix elements. 
To measure the overlap matrix on quantum hardware, we follow the approach described in~\cite{parrish2019}, where one additional ancilla qubit is required, as shown in Fig~\ref{fig:OverlapCircuit}. The price of this simplicity is the requirement for an efficient construction of a controlled time evolution operator. We do this with the use of QSearch~\cite{davis2020towards} for circuit synthesis, where a unitary, $U$, is read in and a circuit with minimal CNOT depth is constructed while keeping the similarity with the synthesized unitary ($U_s$) below some threshold $\sigma$ (here $\sigma = 10^{-10}$ is the Hilbert-Schmidt inner product between the conjugate transpose of $U$ and $U_s$).  
For the 2 qubit transverse field Ising model with one ancilla, a total of 6 CNOTs are needed, regardless of time step~\cite{bassman2021towards}. 
The overlap matrix elements are obtained by measuring $S_{0j} = \langle X \rangle_j + i\langle Y \rangle_j$ on the ancilla qubit for each timestep, where $j$ denotes the number of timesteps after which we make the measurement. As noted in the theory section, because this has the structure of a time correlation function, the other rows of the matrix can be filled out accordingly, with one extra measurement needed for the unitary formulation of the generalized eigenvalue equation. 

\section{Results and Discussion}
\label{section:results}

\subsection{Finding the Optimal Time Step Size}
\label{ss:optDeltat}

\begin{figure*}
    \centering
    \includegraphics[width=1.0\textwidth]{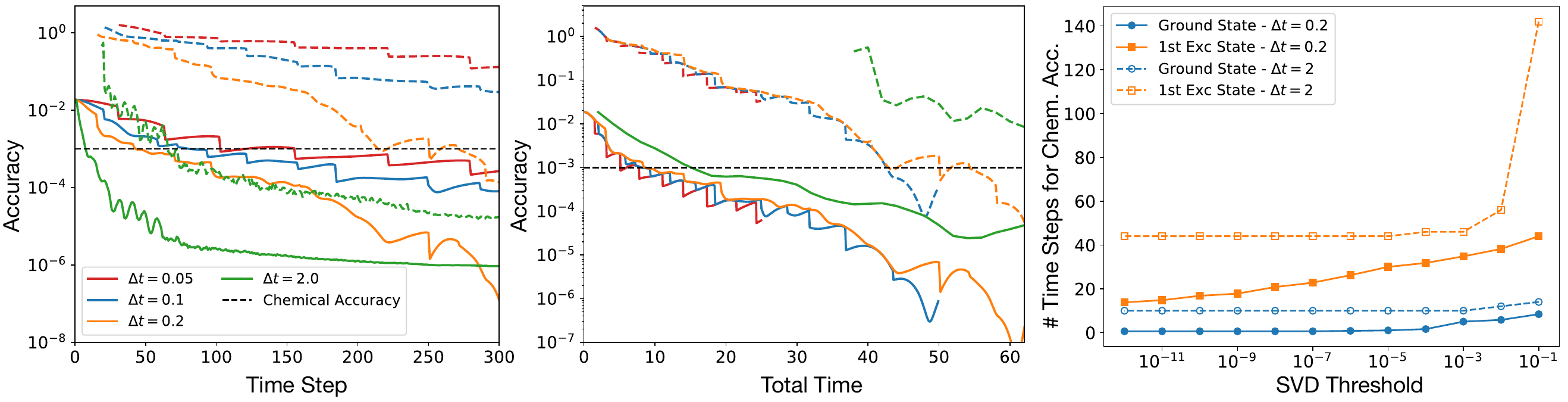}
    \caption{Ground (solid) and excited state (dashed) energy of LiH as a function of timestep (left) and total simulation time (middle).
    The initial reference vector is the Hartree-Fock state. We use the 3-21g basis set which has 4 electrons and 11 orbitals. The SVD cutoff is $10^{-1}$ for both the left and the middle plot and the legend is the same for both.
    The right figure shows the total time to reach chemical accuracy as a function of SVD.}
    \label{fig:lih}
\end{figure*}

Here we exemplify the strategy for finding optimal time step sizes on LiH.
Figure~\ref{fig:lih} shows the convergence of the ground state (solid lines) and first excited state included in the support space (dashed lines) for several time step sizes $\Delta t$, as a function of the number of expansion states (left) and the total simulation time (middle).
The left panel in Fig~\ref{fig:lih} shows how the step-like behavior appearing for the smallest time step size ($\Delta t = 0.05$) decreases with increasing $\Delta t$.
The convergence as a function of simulation time (middle) clearly shows the implicit balance to be made between simulations with short total time and small number of expansion states. 
The $\Delta t = 2.0$ simulation (green curves) achieves chemical accuracy with the smallest number of expansion states, i.e. with the most compact variational Ansatz.
However, this comes at the cost of a significantly larger total simulation time, at least twice as large as the simulations with smaller time step sizes.
For practical implementations on NISQ devices, the marginal decrease in number of expansion states is not worth the drastic increase in total simulation time.
This underlies the importance of finding a time step size that is long enough, but not too long.

We further note that the excited state convergence is worse than the ground state in general, requiring for LiH about twice as many expansion states and total simulation time.
This is likely related to the magnitude of the expansion coefficients of the Hartree-Fock reference vector in the basis of Hamiltonian eigenstates.

The rightmost plot of Fig.~\ref{fig:lih} shows the total simulation time for ground state convergence to chemical accuracy using $\Delta t = 0.2$, for different singular value cutoffs $s_{SV}$. 
Clearly, the larger the value of $s_{SV}$, the longer the simulation time becomes, which relates to the notions of linear independence, which we elaborated upon in the theory section (see Eq.~\eqref{eq:LinDepCond}). 

\subsection{Classical Simulation of Molecular Systems}
\label{ss:class_sim}

\begin{figure*}
    \centering
    \includegraphics[width=\textwidth]{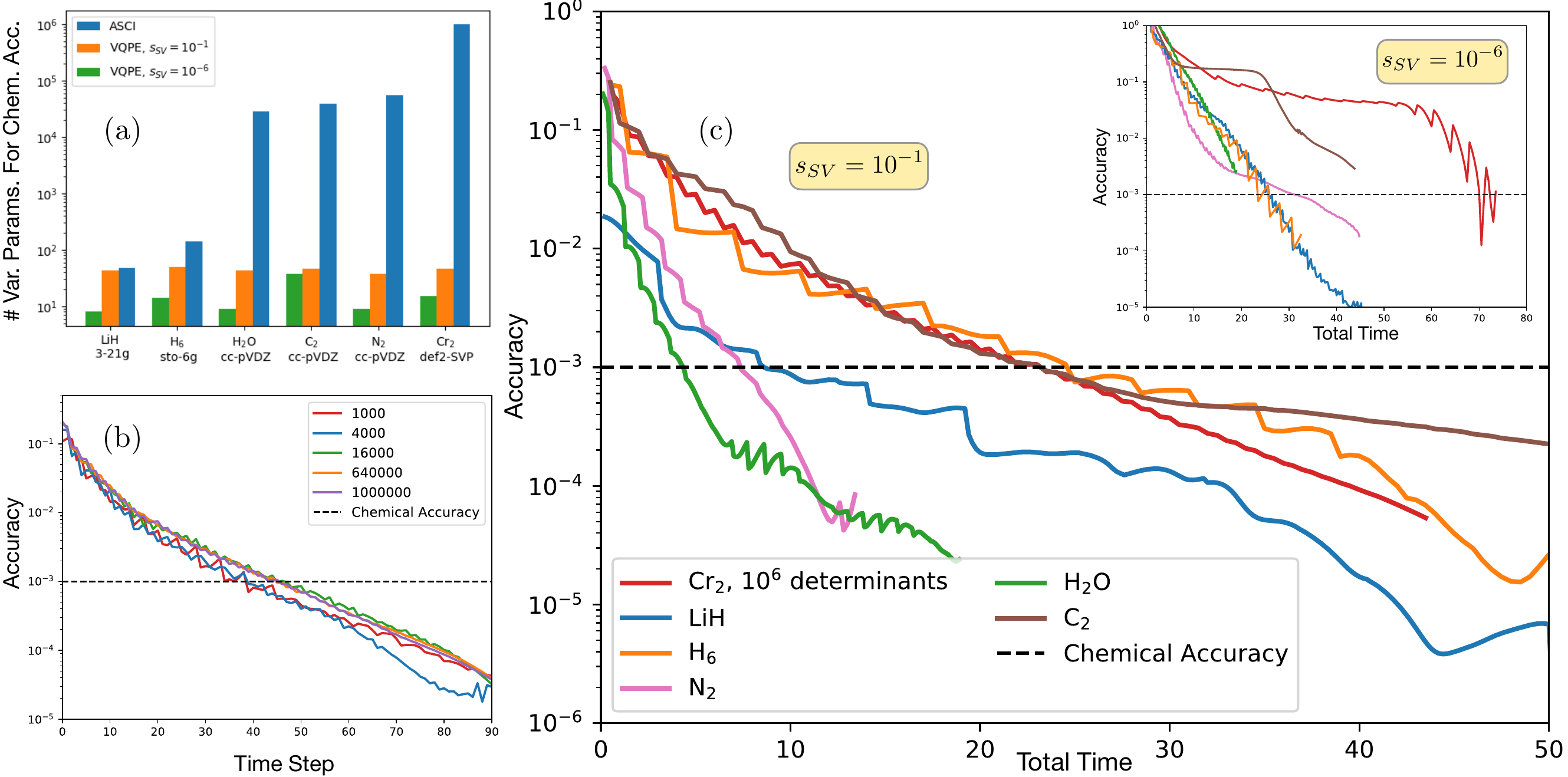}
    \caption{(a) Number of variational parameters to reach chemical accuracy using ASCI (a classical selected configuration interaction approach), and the VQPE algorithm with different $s_{SV}$ values. These parameters correspond to determinants in ASCI, and to expansion states in VQPE. 
    (b) Comparison of convergence for all different Hilbert space sizes of Cr$_2$ (c) Convergence for all the molecules as a function of total time ($s_{SV} = 10^{-1}$) with the inset showing convergence of the first excited state which has  non-zero overlap with the reference state  ($s_{SV} = 10^{-6}$).}
    \label{fig:masterpiece}
\end{figure*}

Figure~\ref{fig:masterpiece} shows the power of the VQPE method on molecular systems with varying degrees of electronic correlation, using a singular value cutoff of $10^{-1}$.
As discussed in~\ref{section:methods}, for all systems except for H$_6$ and LiH, we use a truncated Hamiltonian obtained with ASCI.
In panel (b) we show that this approximation has no effect on our conclusions, since vastly different truncation sizes present the same convergence even though the actual ground state energies are very different. 
Panel (a) in this figure shows the number of variational parameters to reach chemical accuracy using ASCI (a classical selected configuration interaction approach), and the VQPE algorithm with different $s_{SV}$ values.
These parameters correspond to determinants in ASCI, and to expansion states in VQPE.
Given the differences in complexity with respect to the classical algorithm, it is remarkable how many of these systems converge to chemical accuracy at essentially identical rates, as shown in panel (c).
Note that the differences in convergence times are likely connected to the different levels of correlation in these systems: the existence of low lying excited states (i.e. more strongly correlated) means longer convergence time because of a smaller energy gap.
Even with these differences, for all systems studied we reach chemical accuracy with as little as 50 variational parameters and about 30 a.u. of total simulation time.
This exceptionally low number of variational parameters is particularly striking when considering that classically, e.g. with ASCI, it is necessary to include $10^5-10^6$ parameters (determinants) to reach the same energies.

The inset in panel (c)
shows the convergence of the first excited state in the support space.
Here it was necessary to use a significantly smaller SVD cutoff ($s_{SV}=10^{-6}$) to converge in reasonable simulation times.
Even with this drastic restriction in the allowable noise, the excited state energies converge significantly slower than the ground state energies.
Still, they should be possible to access efficiently with reference states tailored to describe excited states~\cite{phillipson1958improved,gilbert2008self,shea2018communication, shea2020generalized,hait2021orbital}.

\subsection{Quantum Simulation of Transverse Field Ising Model}
\label{ss:qsim}

\begin{figure}
    \centering
    \includegraphics[width=0.5\textwidth]{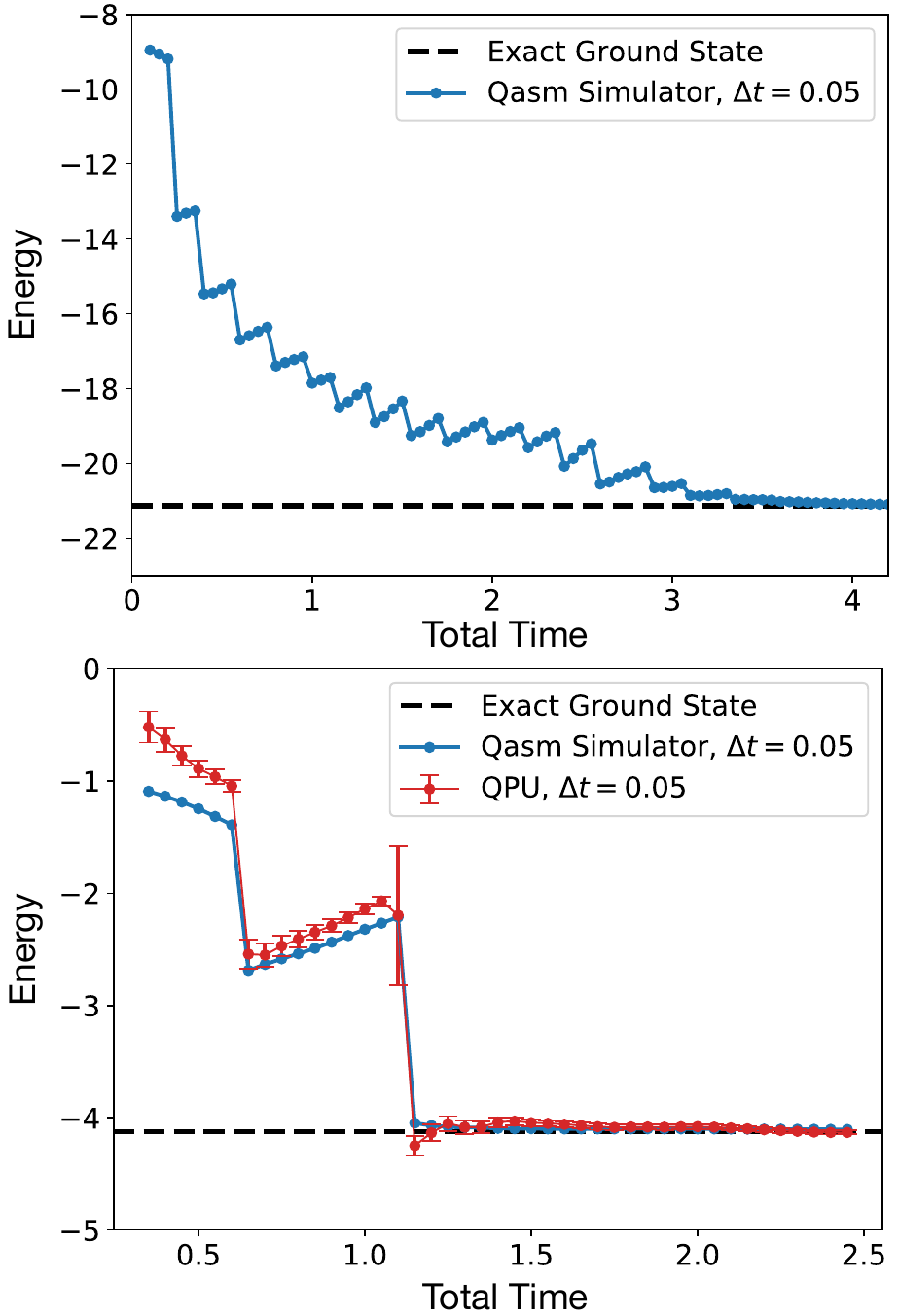}
    \caption{
    The top figure shows the ground state energy as a function of timestep for the 10 qubit transverse field Ising model run on IBM's qasm simulator with $s_{SV} = 10^{-1}$. The bottom figure shows results for hardware runs for a 2 qubit transverse field Ising model. For both the top and bottom plots, VQPE and Trotter step size are the same, $\Delta t = 0.05$. The QPU results were run on IBM's Paris machine. All classical processing in the bottom plot (for both QPU and simulator data) used SVD cutoff $s_{SV} = 2$.}
    \label{fig:hardware}
\end{figure}

Finally, Fig.~\ref{fig:hardware} shows our results using IBM's QPUs (2 qubit TFIM) and qasm simulator (10 qubit TFIM), compared to the exact ground state energy. Our initial state corresponds to a ferromagnetically ordered chain where all spins point down in the $Z$ direction, which corresponds to the $|000...\rangle$ state in the computational basis.
For processing the hardware data we use $s_{SV} = 2$ to account for larger noise.
Both the simulator and hardware runs used the same Trotter and VQPE step size $\Delta t = 0.05$.
Error bars on the hardware data correspond to the variance of the average ground state energy at each time point over 10 individual hardware runs on IBM's Paris machine, using qubits 0, 1, and 2. 
Each hardware run used 8192 shots per matrix element.
Figure~\ref{fig:hardware} shows convergence to the exact ground state energy for both simulator and hardware data.
The saw-tooth behavior is indicative of linear dependence in the expansion states, as discussed and shown in Sections~\ref{ss:pcc} and~\ref{ss:optDeltat}.

\section{Conclusions}
\label{section:conclusion}

Here we analyze the theoretical underpinnings of VQPE algorithms as well as the effects of statistical noise on its performance.
We have presented evidence for VQPE as a particularly compact and natural algorithm for the NISQ era, showing that with the unitary formulation of the generalized eigenvalue equation and the prevailing Toeplitz structure of the overlap matrix, only a linear number of measurements are needed.
We provided a heuristic for choosing an optimal time step size, which balances two opposing factors: On the one hand, generating linearly independent expansion states with each new time evolution, such that the variational Ansatz becomes compact, which sets a lower bound to the time step size, and 
on the other hand minimizing the total simulation time, as required by NISQ hardware.
We have also demonstrated the effect of statistical noise on the optimal time step and final accuracy of the energies, showing that simple regularization techniques suffice to mitigate the effects of noise.

We have exemplified the power of the VQPE approach on a wide range of molecules of different complexities, simulating the algorithm classically, as well as the transverse field Ising model on IBM's quantum simulator and hardware.
On a traditional example of strong correlation, Cr$_2$, our results suggest that VQPE achieves comparable accuracy to state of the art classical simulations with orders of magnitude fewer variational parameters ($\sim50$ vs. $10^6$). 
This compactness, together with their NISQ compatibility (with respect to total simulation time, number of measurements, and noise resilience), marks VQPE approaches as some of the most promising platforms to realize the long sought quantum computation goal of performing many-body simulations beyond the reach of classical computations.

\section*{Acknowledgments}
K.K., M.U., and W.A.d.J. were supported by the the Office of Science, Office of Advanced Scientific Computing Research Quantum Algorithms Team Programs, and C.M.Z. was supported by Basic Energy Sciences, Chemical Sciences, Geosciences and
Biosciences Division, as part of the Computational Chemical Sciences Program, of the U.S. Department of Energy under Contract No. DE-AC02-05CH11231. N.M.T., S.C. and F.W. are grateful for support from NASA Ames Research Center and support from the AFRL Information Directorate under Grant No.~F4HBKC4162G001. F.W. was supported by NASA Academic Mission Services, contract number NNA16BD14C.
Some of the calculations were performed as part of the XSEDE computational Project No. TG-MCA93S030. We acknowledge the use of the Qiskit~\cite{aleksandrowicz2019qiskit} software package for performing quantum simulations. This research used
resources of the Oak Ridge Leadership Computing Facility, which is a
DOE Office of Science User Facility supported under Contract
No.~DE-AC05-00OR22725.
NW was funded by a grant from Google Quantum AI, and his theoretical work on bounding simulation costs was  supported by the U.S. Department of Energy, Office of Science, National Quantum Information Science Research Centers, Co-Design Center for Quantum Advantage under contract number DE-SC0012704.
We thank William J. Huggins and Andr\'{e}s Montoya-Castillo for insightful discussions. We thank Jonathan S. Landy for a careful reading of the manuscript and helpful suggestions.

%

\end{document}